\documentclass[10pt, conference, compsocconf]{IEEEtran}
\usepackage[letterpaper=true,pdfview=FitH,pdfstartview=FitH,bookmarks=true]{hyperref}

\pdfoutput=1

\newcommand{\deanonypaper}{DEANONYPAPER} 
\newcommand{\fullpaper}{FULLPAPER} 

\usepackage{cite}

\ifCLASSINFOpdf
\else
\fi
\usepackage[cmex10]{amsmath}
\usepackage{amsfonts}                 
\usepackage{amssymb}                  
\usepackage{latexsym}                 
\usepackage{graphicx}                 
\usepackage{url}                      
\usepackage{ifthen}

\newcommand{\term}[1]{\emph{#1}}

\newcommand{\eg}[0]{\textit{e.g.,}}   
\newcommand{\ie}[0]{\textit{i.e.,}}   
\newcommand{\cf}[0]{\textit{cf.}}     
\newcommand{\Eg}[0]{\textit{E.g.,}}   
\newcommand{\MC}[0]{\texttt{mozilla-central}} 
\newcommand{\BZ}[0]{\texttt{Bugzilla}} 
\newcommand{\LS}[0]{\texttt{libsvm}}  
\newcommand{\nth}[1]{\ensuremath{^{\mbox{\scriptsize{#1}}}}}
\renewcommand{\Pr}[1]{\mathrm{Pr}\left({#1}\right)}
\newcommand{\Exp}[1]{\mathbf{E}\left[{#1}\right]}
\newcommand{\Gain}[1]{\mathrm{Gain}(#1)}
\newcommand{\GainRatio}[1]{\mathrm{GainRatio}(#1)}
\newcommand{\Ent}[1]{\mathrm{Entropy}(#1)}
\newcommand{\eref}[1]{(\ref{#1})}

\renewcommand{\paragraph}[1]{\par \noindent \textbf{{#1}}}
\newtheorem{theorem}{Theorem}[section]
\newtheorem{remark}[theorem]{Remark}

\begin{document}
\ifthenelse{\isundefined{\deanonypaper}}{}{\IEEEoverridecommandlockouts}

%
\title{How Open Should Open Source Be?}


\ifthenelse{\isundefined{\deanonypaper}}{}{
\author{\IEEEauthorblockN{Adam Barth}
\IEEEauthorblockA{Google\\
adam@adambarth.com}
\and
\IEEEauthorblockN{Saung Li}
\IEEEauthorblockA{CS Division, UC Berkeley\\
shadowcwal@berkeley.edu}
\and
\IEEEauthorblockN{Benjamin I. P. Rubinstein}
\IEEEauthorblockA{Microsoft Research\\
ben.rubinstein@microsoft.com}
\and
\IEEEauthorblockN{Dawn Song}
\IEEEauthorblockA{CS Division, UC Berkeley\\
dawnsong@eecs.berkeley.edu}
}}


%


\maketitle

\begin{abstract}
Many open-source projects land security fixes in public repositories before
shipping these patches to users. This paper presents attacks on such 
projects---taking Firefox as a case-study---that exploit patch metadata
to efficiently search for security patches prior to shipping.
Using access-restricted bug reports linked from patch descriptions,
security patches can be immediately identified for 260 out of 300 days
of Firefox 3 development. In response to Mozilla
obfuscating descriptions, we show that machine learning can exploit
metadata such as patch author to search for security patches, extending
the total window of vulnerability by 5 months in an 8 month period when
examining up to two patches daily. Finally we present strong evidence that 
further metadata obfuscation is unlikely to prevent information leaks,
and we argue that open-source projects instead ought to keep security
patches secret until they are ready to be released.
\end{abstract}


%
\IEEEpeerreviewmaketitle

\section{Introduction}\label{sec:intro}

Many open-source software development projects, such as Firefox, Chromium,
Apache, the Linux kernel, and OpenSSL, 
produce software that is run by hundreds of millions of users and
machines. Following the open-source spirit, these projects make all code
changes immediately visible to the public in open code repositories, including
landing fixes to security vulnerabilities in public development
branches before publicly announcing the vulnerability and providing an updated
version to end users. This common practice raises the question of whether this
extreme openness increases the \term{window of vulnerability} by enabling
attackers to discover vulnerabilities earlier in the security life-cycle, \eg\ via
program analysis~\cite{autoexploit}. The conventional wisdom is
that detecting these security patches is made prohibitively difficult because the
patches are hidden among a cacophony of non-security changes.
For example, the central Firefox repository receives, on average, $38.6$ patches
per day, of which $0.34$ fix security vulnerabilities. Recently, 
blackhats in the Metasploit project have used the ``description'' metadata
field to find Firefox patches that refer to non-public bug numbers~\cite{dveditz}. The
Firefox developers have responded by obfuscating the description field, but
where does this cat-and-mouse game end?

In this paper, we analyze information leaks in open-source life-cycles, through a
case-study on Firefox~3 and~3.5, to answer three key questions: (1) Does the
metadata associated with patches in the source code repository contain
information about whether the patch is security sensitive? (2) Using this
information as a guide, how much less effort does an attacker need to expend
to find unannounced security vulnerabilities? (3) By how much do these information
leaks increase the total window of vulnerability? We do not study the task of 
reverse engineering exploits from patches~\cite{autoexploit}, but instead
focus on prioritizing the search for security patches to decrease attack cost.
In so doing we make several contributions:
\begin{itemize}
\item For all but 40 out of 300 days of Firefox 3 active development, 
a simple join of Firefox's repository and bug tracker identifies a newly landed 
security patch.
\item Even if the patch-bug ID link is obfuscated perfectly, off-the-shelf
machine learning can rank patches using remaining metadata such that
examining up to two patches daily will add an extra 5 months of vulnerability within
an 8 month period of Firefox 3 development.
\item We offer strong evidence that obfuscating patch metadata achieves
diminishing returns, and even under complete obfuscation a random
ranker adds over 2 months of vulnerability by examining just 2 patches daily.
\item We argue that instead of obfuscating metadata, open-source projects
like Firefox should land security patches in a private release branch accessible to
trusted testers.
\end{itemize}

We first quantify the severity of information leaks due to Firefox's bug tracker \BZ.
Upon arrival of a new patch
in the \MC\ main development trunk, it is a simple matter for an attacker to parse
the patch's description field for a bug ID and then consult \BZ\ for the corresponding
bug history. We show that for the vast majority of days of active Firefox development,
a corresponding \BZ\ bug report page was access restricted at the time of landing,
divulging the patch's nature as fixing a vulnerability, in which case the attacker need
only reverse-engineer one patch to find a zero-day vulnerability.

Prompted by Metasploit blackhats exploiting this information leak,
Mozilla recently began obfuscating patch descriptions~\cite{dveditz}.
Given the wealth of metadata still available, including
patch author, the set of files modified, and the size of these modifications, 
we conjectured that machine learning could significantly narrow 
an attacker's search space of patches. 
While we show that each feature individually contains little information about
whether patches are security sensitive, our experiments on Firefox 
show that a non-linear combination of features provides valuable guidance for 
attackers.  Our attack uses an off-the-shelf implementation of
a support vector machine~(SVM) trained to discriminate between security
and non-security patches based on non-description metadata. We then
use the SVM to rank newly-landed patches by likelihood of fixing a 
vulnerability.
We measure the cost to the adversary by
\term{attacker effort}---the number of patches the attacker would need to
examine before finding the first vulnerability; for over a third of an 8 month
period of Firefox 3, an SVM-assisted attacker discovers a security patch upon the first
patch examined daily. We measure the benefit to the attacker by the
\term{increase to the window of vulnerability}; an attacker who examines the top
two patches ranked by the detection algorithm each day will add an extra $148$ days
of vulnerability to the $229$ day period we study, representing a 6.4-fold increase
over the window of vulnerability caused by the latency in deploying security updates.

In performing the cost-benefit analysis of the SVM ranker, we also consider
a random ranker which searches through recently landed patches in random order.
While the random ranker serves as a benchmark with which to judge the SVM's 
performance, it demonstrates that even under perfect obfuscation of \emph{all}
patch metadata, an attacker need not expend significant effort to yield moderate results.

Our results suggest that Firefox should change its security life-cycle to
avoid leaking information about unannounced vulnerabilities in its public
source code repositories. Instead of landing security patches in the central
repository, Firefox developers should land security patches in a private
release branch that is available only to a set of trusted testers. The
developers can then merge the patches into the public repository at the same
time they release the security update to all users and announce the
vulnerability.

Although we study Firefox specifically, we believe our results
generalize to other open-source projects, including
Chromium, Apache, the Linux kernel, and OpenSSL, which land vulnerability
fixes in public repositories before announcing the vulnerability and
making security updates available. However, we choose to study these issues in
Firefox because Firefox has a state-of-the-art process for responding
to vulnerability reports and publishes the ground truth about which patches fix security vulnerabilities~\cite{known-vulnerabilities}.

\paragraph{Organization.}
The remainder of the paper is organized as follows.
Section~\ref{sec:lifecycle} describes the existing Firefox security
life-cycle. Section~\ref{sec:analysis-goals} lays out the dataset we analyze.
Section~\ref{sec:methodology} explains our methodology.
Section~\ref{sec:results} presents our results.
Section~\ref{sec:secure-lifecyle} recommends a secure security life-cycle.
Section~\ref{sec:conclusions} concludes.

\section{Life-Cycle of a Vulnerability}\label{sec:lifecycle}

This section describes the life-cycle of a security patch for Firefox. We take Firefox as a representative example, but many open-source
projects use a similar life-cycle.

\subsection{Stages in the Life-Cycle}

\begin{figure}[t]
\begin{center}
\begin{minipage}[t]{1\columnwidth}
\centering
\includegraphics[width=1\linewidth]{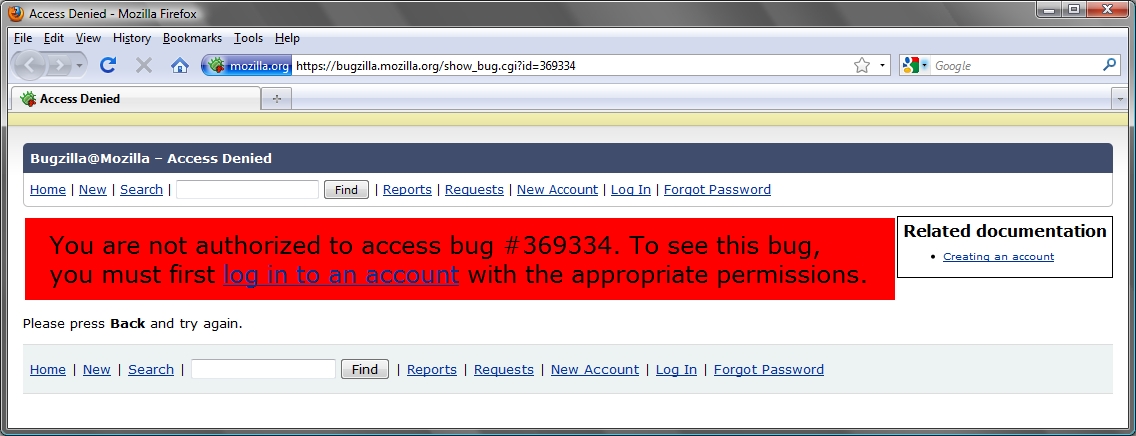}
\caption{Information leaked about security-sensitive bug numbers has 
previously been exploited by attackers to identify undisclosed vulnerabilities,
when bug numbers were linked from landed patches via patch descriptions.}
\label{fig:denied}
\end{minipage}
\end{center}
\end{figure}

In the Firefox open-source project, vulnerabilities proceed through a sequence of observable stages:
\begin{enumerate}
\item \label{stage:filed} \textbf{Bug filed.} The Firefox project encourages security researchers
to report vulnerabilities via the project's public bug tracker \BZ.
When filed, security bugs are marked ``private" (meaning access is restricted
to a trusted set of individuals on the security
team~\cite{mozilla-security-group}; see Figure~\ref{fig:denied}) and are assigned a
unique ID.

\item \label{stage:landed} \textbf{Patch landed in mozilla-central.} Once the developers determine
the best way to fix the vulnerability, a developer writes a patch for the
mainline ``trunk'' of Firefox development. Other developers review the patch
for correctness. Once the patch is approved, the developer \term{lands}
the patch in the public \MC\ Mercurial repository.

\item \textbf{Patch landed in release branches.} After the patch successfully
lands on \MC\ (including passing all the automated regression and performance
tests), the developers \term{merge} the patch to one or more of the Firefox
release branches.

\item \label{stage:released} \textbf{Security update released.}
At some point, a release driver decides to release an updated version of
Firefox containing one or more security fixes (and possibly some non-security
related changes). These releases are typically made from the release branch,
not from the \MC\ repository. The current state of the release branch is
packaged, signed, and made available to users via Firefox's auto-update
system. 

\begin{figure*}[t]
\begin{center}
\begin{minipage}[t]{1\textwidth}
\centering
\includegraphics[width=1\textwidth]{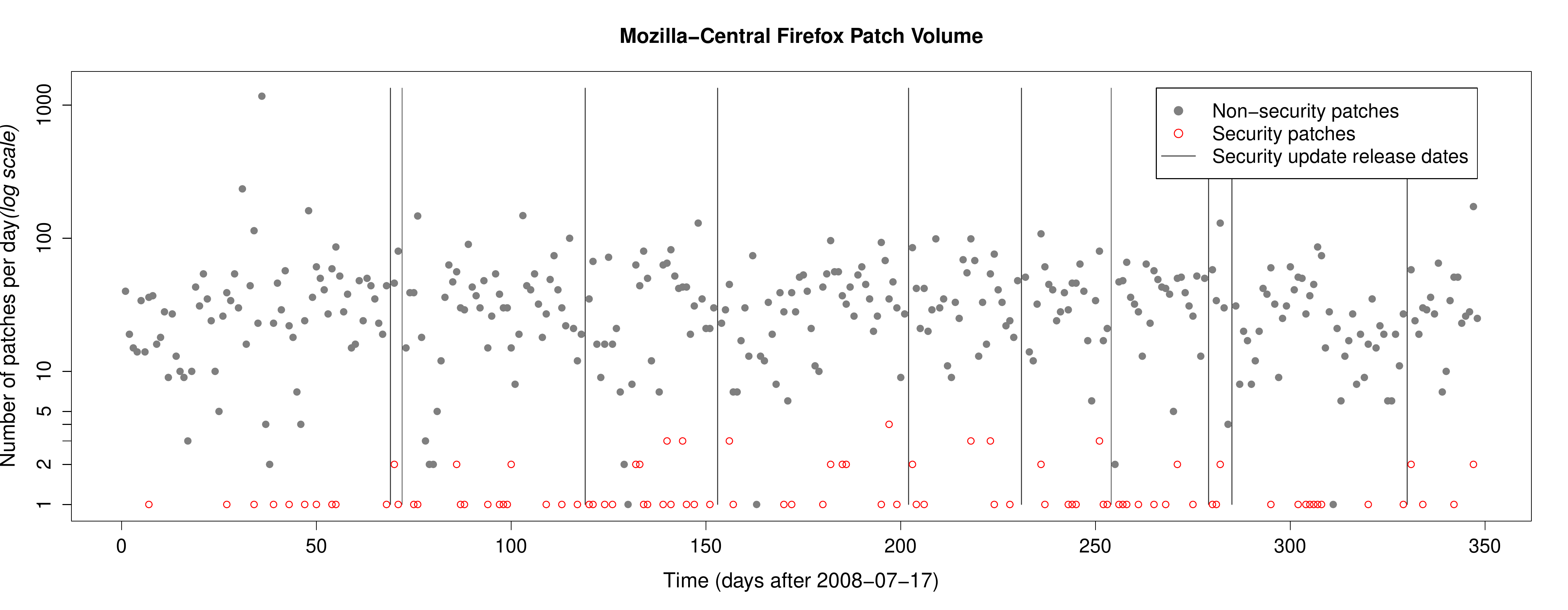}
\caption{Attackers must find security patches within a ``thundering herd'' of
non-security patches.}
\label{fig:thundering-herd}
\end{minipage}
\end{center}
\end{figure*}

\item \textbf{Vulnerability announced.} Mozilla announces the
vulnerabilities fixed in the release~\cite{known-vulnerabilities}. For the
majority of vulnerabilities, disclosure is simultaneous with the release. However, in some
cases disclosure can occur weeks later (after the security update is applied by most users).

\item \label{stage:applied} \textbf{Security update applied.} Once a user's auto-update client
receives an updated version of the Firefox binary and the user chooses to 
install the binary, Firefox updates itself. Upon installation, the user is
protected from an attacker exploiting the vulnerability.
\end{enumerate}
Previous work~\cite{update-dynamics} has analyzed the dynamics between
steps (\ref{stage:released}) and (\ref{stage:applied}), finding that the user experience and download
size have a dramatic effect on the time delay and, hence, the window of
vulnerability.  With a sufficiently well-designed update experience, browser
vendors can reduce the lag between (\ref{stage:released}) and (\ref{stage:applied}) to a matter of days. Recent
releases of Firefox have an improved update experience that reduces the window
of vulnerability between steps (\ref{stage:released}) and (\ref{stage:applied}).

However, not as much attention has been paid to the dynamics between steps (\ref{stage:filed}) and
(\ref{stage:released}), likely because most people make the assumption that little is
revealed about a vulnerability until the vulnerability is intentionally
disclosed in step (\ref{stage:released}). Unfortunately, there are a number of information leaks
in this process that invalidate that assumption.

\subsection{Information Leaks in Each Stage}

Each stage in the vulnerability life-cycle leaks some amount of information
about vulnerabilities to potential attackers. For example, even step~(\ref{stage:filed})
leaks information because bug numbers are issued sequentially
and a brute-force attack can determine which are
``forbidden'' and hence represent security vulnerabilities. Of course, simply
knowing that a vulnerability was reported to Firefox does not give an
attacker much useful information for creating an exploit.

More information is leaked in stage~(\ref{stage:landed}) when developers land security patches in \MC\
because \MC\ is a public repository. It is unclear, \emph{a priori},
whether an attacker will be able to find security patches landing in \MC\
because these security patches are landed amid a ``thundering herd'' of other
patches (\cf\ Figure~\ref{fig:thundering-herd}), but if an attacker can detect
that a patch fixes a security vulnerability, the attacker can learn
information about that vulnerability. For example, the attacker learns where in
the code base the vulnerability exists. If the patch fixes a vulnerability by
adding a bounds check, the attacker can look for program inputs that generate
large buffers of the checked type. In this paper, we do not evaluate the
difficulty of reverse engineering an exploit from a vulnerability fix, but
there has been some previous work~\cite{autoexploit} on reverse engineering
exploits from binary patches (which is, of course, more difficult than reverse
engineering exploits from source patches).  Instead 
we focus on reducing attacker effort through efficient methods for prioritizing
their search.

\section{Analysis Goals and Setup} \label{sec:analysis-goals}

In this section, we describe the dataset and the performance metrics for searching for security patches.

\subsection{Dataset}

\paragraph{Set of Patches.} In our experiment, we focused on the complete
life-cycle of Firefox~3, which lasted over $12$ months, contained $14,\!416$
non-security patches, $125$ security patches, and $12$ security updates. 
We use publicly available data starting from the release of
Firefox~3 and ending with the release of Firefox~3.5. Also, to strengthen our
results, we focus on the \MC\ repository, which receives the vast majority of
Firefox development effort. We cloned the entire \MC\ repository to our
experimental machines to identify all patches during the life-cycle of
Firefox~3. We ignore the release branches to evaluate how well our search
methods are able to find security fixes amid mainline development (see
Figure~\ref{fig:thundering-herd}).

Although we focus our attention to Firefox~3, we repeat our results on
Firefox~3.5 and expect our results to generalize to other releases of Firefox and
to other open-source projects.
Firefox 3.5 was released June 30, 2009 and remained active until the release of
Firefox 3.6 on January 21, 2010. During this 6 month period, 7 minor releases 
to Firefox were made ending with Firefox 3.5.7 on January 5, 2010.
During the 6 month period, 7,033
patches were landed of which 54 fixed vulnerabilities.

\paragraph{Ground Truth.} We determined the ``ground truth'' of whether a patch fixes a security
vulnerability by examining the list of known vulnerabilities published
by Firefox~\cite{known-vulnerabilities}.  Each list of Common Vulnerabilities
and Exposures on the known
vulnerability web page contains a link to one or more entries in \BZ. 
At the time we crawled these bug entries (after disclosure),
the entries were public and contained links to the Mercurial commits
that fixed the vulnerabilities (both in \MC\ and in the release branches).
Our crawler harvested these links and extracted the unique identifier for
each patch. 

The known vulnerability page dates each vulnerability disclosure, and we
assume that these disclosure dates are accurate. Each bug entry is timestamped
with its creation date and every message on the bug thread is dated as well.
Finally, the \MC\ pushlog website contains the date and time of every change in
the ``pushlog,'' which we also assume is authoritative.

\subsection{Attack Performance Metrics}

We consider attackers who search for security patches in open-source repositories,
with the goal of reverse-engineering them to produce zero-day exploits.
We focus on three different methods for ranking patches for an attacker to search
through, to find a security vulnerability. Each search method can be thought
of as producing a ranking on the pool of recently landed patches, where better
rankings place (at least) one security patch higher than non-security patches. 
The attacker examines the landed patches in rank order, until a security patch is
found. The ranker's usefulness, as formalized below, lies in the reduction to attacker
effort and in the increase to the window of vulnerability.

\subsubsection{Cost of Vulnerability Discovery: Attacker Effort} \label{sec:attacker-effort}

Given a set of patches and a ranking function, we call the rank of the first
true security patch the \textit{attacker effort}. This quantity reflects the
number of patches the attacker has to examine when searching the ranked list
before finding the first patch that fixes a security vulnerability. For
example, if the third-highest ranked patch actually fixes a security
vulnerability, then the attacker needs to examine three patches before finding the
vulnerability, resulting in an attacker effort of three. Using this metric, we
can compute the percent of days on which an attacker who expends a given
effort will be able to find a security patch.

\subsubsection{Benefit for the Attacker: Window of Vulnerability}\label{subsec:vulnerability-window}

Another metric we propose is the \term{increase to the window of
vulnerability} due to the assistance of a ranker. In
particular, 
an attacker who discovers a vulnerability $d$ days
before the next security update increases the total window of vulnerability
for Firefox users by $d$ days. (Notice that knowing of multiple vulnerabilities
simultaneously does not increase the aggregate window of vulnerability because
knowing multiple vulnerabilities simultaneously is redundant.)

Previous work~\cite{Duebendorfer} explores the effectiveness of browser update
mechanisms, finding that security updates take some amount of time to
propagate to users. In particular, they measured the cumulative distribution
of the number of days users take to update their browsers after security
updates~\cite[Figure~3]{Duebendorfer}. After about $10$ days, the penetration
growth rate rapidly decreases, asymptotically approaching about $80\%$. By
integrating the area above the CDF up to $80\%$, we can estimate the expected
number of days a user takes to update Firefox~3 conditioned that they are in
the first $80\%$ who update. We estimate this quantity, the
\term{post-release window of vulnerability}, to be $3.4$ days, which we
use as a baseline value for comparing windows of vulnerability.

\section{Methodology} \label{sec:methodology}

In this section, we describe the methodology we use to analyze information
leaks in the open-source security life-cycle. We present three security patch search methods,
by order of decreasing levels of patch metadata used.

\subsection{Exploiting Patch Description and Bug History}
\label{sec:bugalgo}

\begin{figure}[t]
\begin{center}
\begin{minipage}[t]{1\columnwidth}
\centering
\includegraphics[width=1\columnwidth]{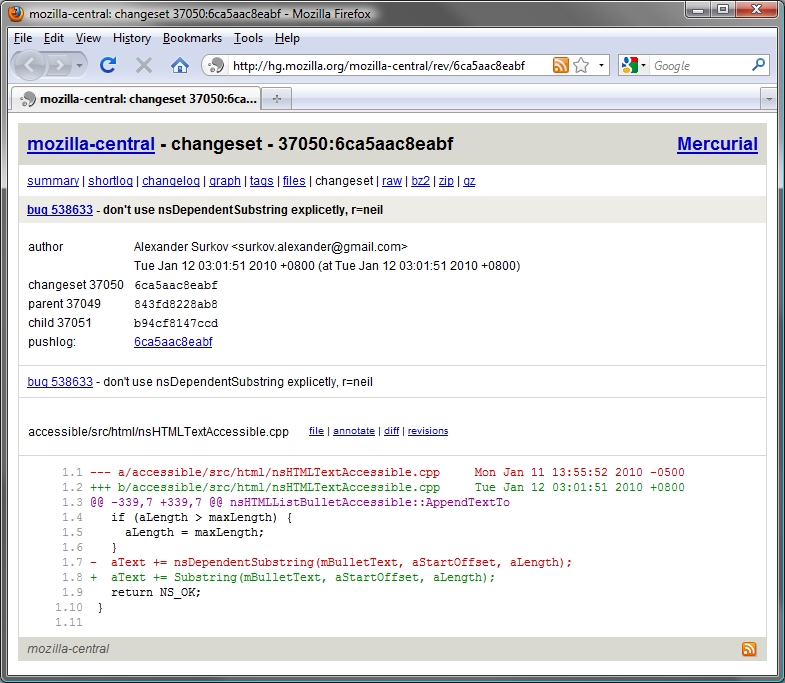}
\caption{An example patch with metadata from the Firefox Mercurial repository.
In addition to patch description and bug number, several features leak
information about the security-related nature of a patch.}
\label{fig:mercurial}
\end{minipage}
\end{center}
\end{figure}

Our simplest approach to discovering security patches is an attack recently
observed in the wild~\cite{dveditz}. \MC\ metadata can include a patch 
description which regularly references the number of the bug fixed
by the patch (\cf\ Figure~\ref{fig:mercurial}). When a new patch lands, the
attacker can parse the description field for bug numbers, and query 
\BZ\ for the corresponding bug history. If the bug history page is access
restricted (\cf\ Figure~\ref{fig:denied}) or it records a previous
core-security add/remove event (a flag indicating that access to the bug report
is restricted to the security team), then the attacker can be confident that
the landed patch fixes a vulnerability. After crawling all patches in \MC\ we
simulated this attack by parsing description fields and querying \BZ; if an
add/remove event occurred prior to the simulated day of examination then
we infer that
an attacker would have correctly deemed the patch to be security-related.

\subsection{A Learning-Based Ranker} \label{sec:svmalgo}

Given a collection of past patches, an attacker can label them based on
whether the patches have been announced as vulnerability fixes. Using these
labels, the attacker can train a statistical machine learning algorithm to
predict whether a new patch fixes an unannounced vulnerability---a 
supervised binary classification problem.
We consider learners that output a real-valued confidence for their
predictions. The attacker can
use these confidence values to rank a set of patches, before examining
the patches in rank order.

\paragraph{Features used by the Learner.}
There are a number of features we could use to identify security patches.
When Mozilla became aware of the previous attack, they began to take steps
to obfuscate patch descriptions~\cite{dveditz}.
Obfuscating and de-obfuscating the patch description is clearly a
cat-and-mouse game. To simulate perfect obfuscation of the patch description by the Firefox developers, we analyze information leaks in other metadata associated with each patch (\cf\ Figure~\ref{fig:mercurial}).
\begin{itemize}
\item \textbf{Author.} We hypothesize that information about the patch author
(the developer who wrote the patch) will leak a sizable amount of information
because Firefox has a security team~\cite{mozilla-security-group} that is
responsible for fixing security vulnerabilities. Most members of the Firefox
community do not have access to security bugs and are unlikely to write
security patches.

\item \textbf{Top-level directory.} For each file that was modified by the
patch, we observe the top-level directory in the repository that contained
the file. In the Firefox directory structure, the top-level directory roughly
corresponds to the module containing the file. If a patch touches more than
one top-level directory, we pick the directory that contains the most
modified files.

\item \textbf{File type.} For each file that a patch modified, we
observe the file's extension to impute the file's type. \Eg\ Firefox patches
often modify C++ implementation files, interface
description files, and XML user interface descriptions. If a patch touches
more than one type of file, we pick the type with the most modified
files.

\item \textbf{Patch size.} We observe a number of size metrics for each patch,
including the total size of the diff in characters, the number of lines in the
diff, the number of files in the diff, and the average size of all modified
files. Although these features are highly correlated, the SVM's ability to
model non-linear patterns lets us take advantage of all these features.

\item \textbf{Temporal.} The timestamp for each patch reveals the time of day
and the day of week the patch was landed in the \MC\ repository. We include
these features in case, for example, some developers prefer to land security
fixes at night or on the weekends.
\end{itemize}

\ifthenelse{\isundefined{\fullpaper}}{}{
\begin{figure*}[t]
\begin{center}
\begin{minipage}[t]{0.48\textwidth}
\centering
\includegraphics[width=1\linewidth]{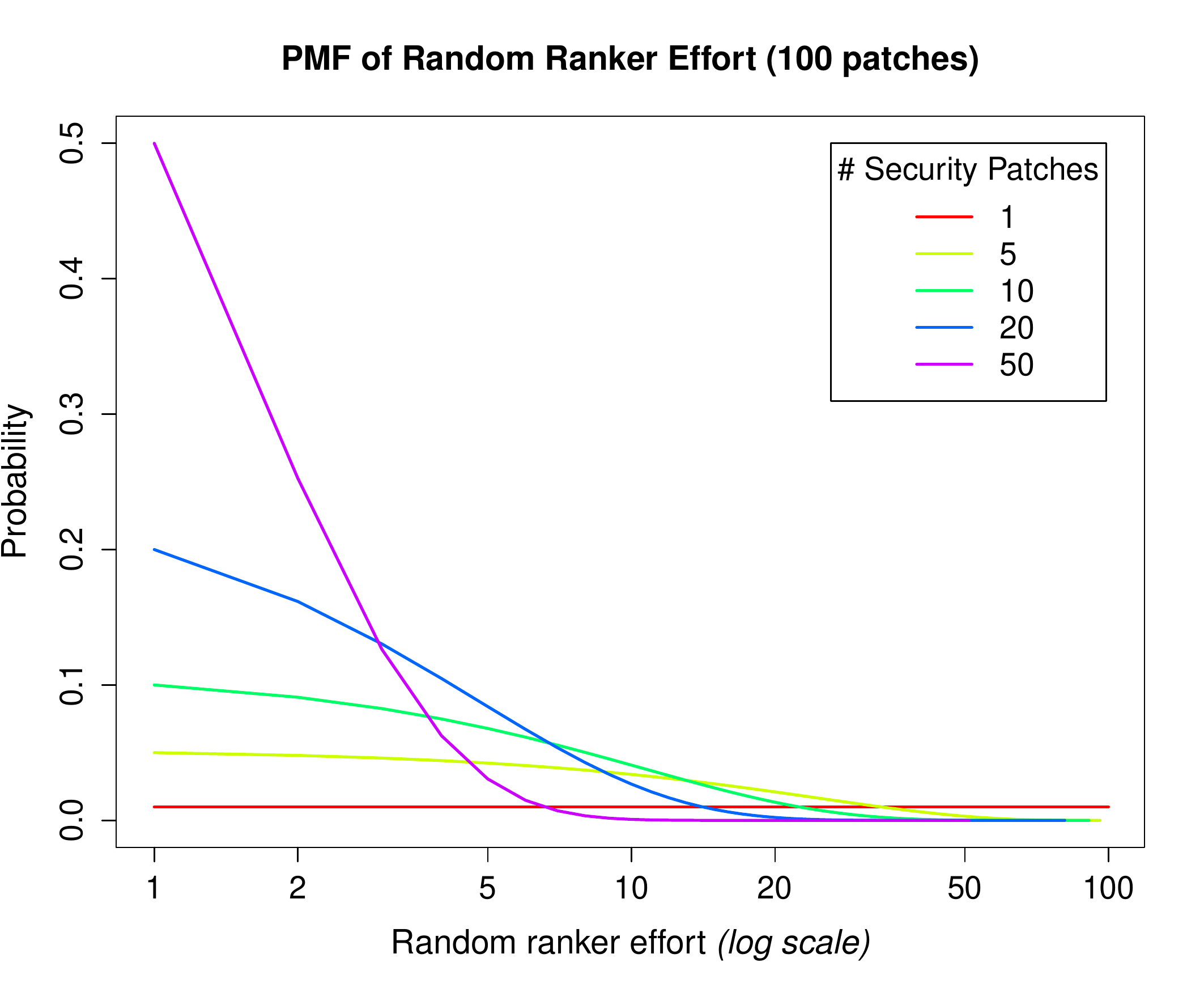}
\ifthenelse{\isundefined{\fullpaper}}{
\caption{The distribution of the random ranker's effort $X$ as a function of $n_s$ for $n=100$.}
}{
\caption{The distribution of the random ranker's effort $X$ as a function of $n_s$ for $n=100$, as given by Equation~\eqref{eq:random-prob}.}
}
\label{fig:random-prob}
\end{minipage}\hfill
\begin{minipage}[t]{.48\textwidth}
\centering
\includegraphics[width=1\linewidth]{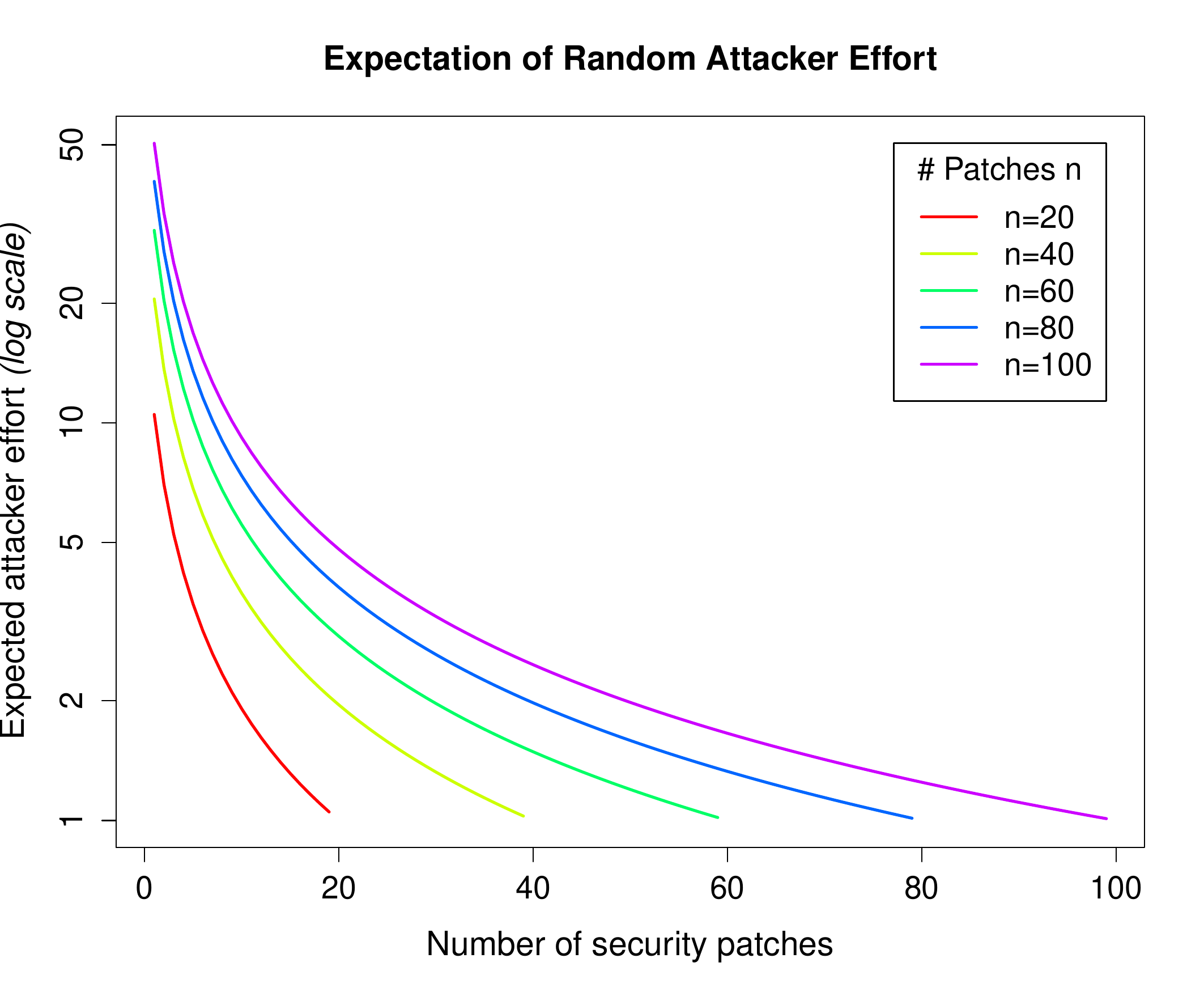}
\caption{The random ranker's expected effort $\Exp{X}$ as a function of 
$n_s$ for $n\in\{20,40,60,80,100\}$, as given by Equation~\eref{eq:random-exp}.}
\label{fig:random-exp}
\end{minipage}
\end{center}
\end{figure*}
} 

\begin{figure*}[t]
\begin{center}
\begin{minipage}[t]{.48\textwidth}
\centering
\includegraphics[width=1\linewidth]{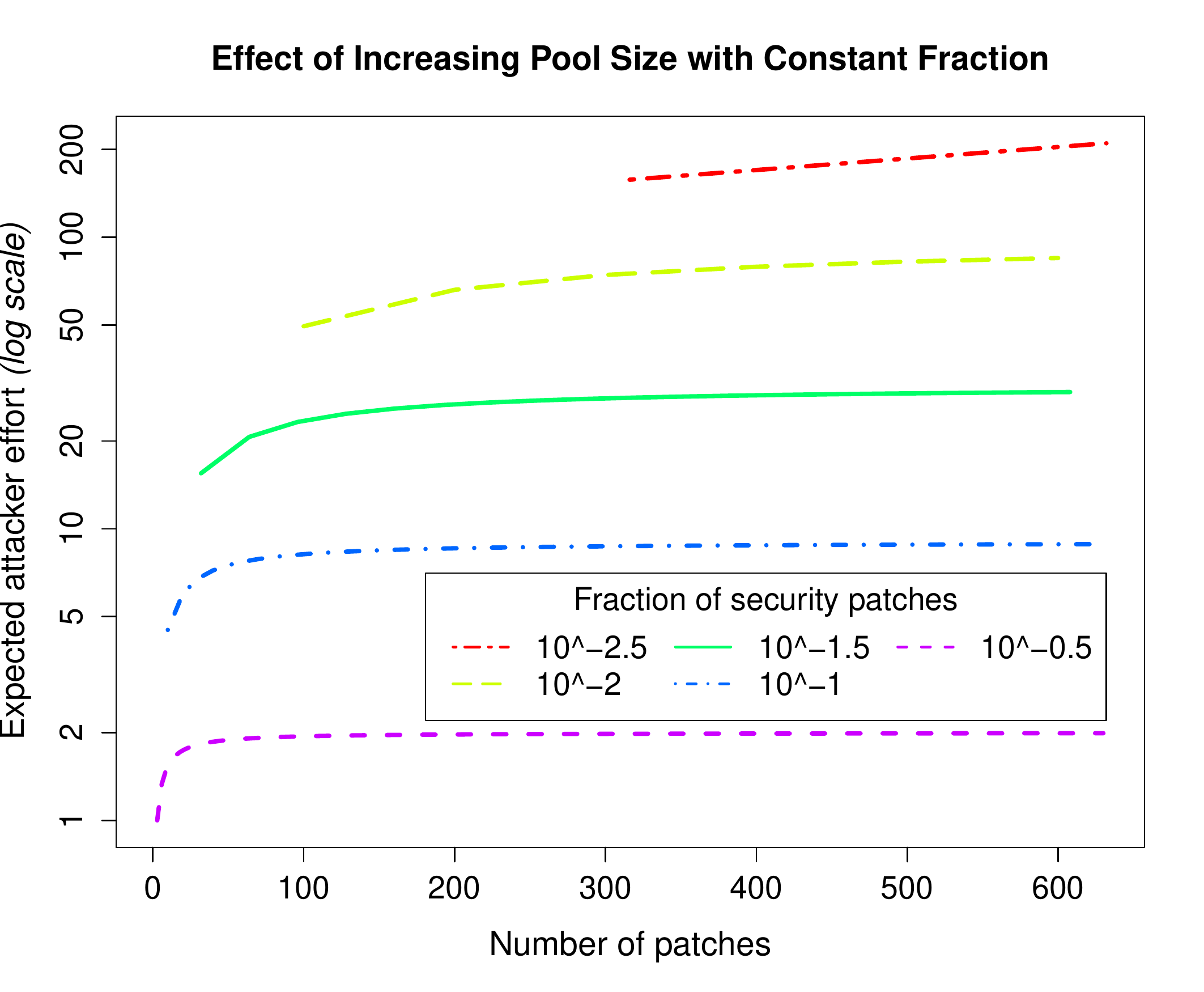}
\caption{The random ranker's expected effort $\Exp{X}$ as a function of 
$n$ for constant fractions of security patches $n_s/n\in\{0.0032,0.01,0.032,0.1,0.32\}$, as given by Equation~\eref{eq:random-exp}.}
\label{fig:random-const-frac-eff}
\end{minipage}\hfill
\begin{minipage}[t]{.48\textwidth}
\centering
\includegraphics[width=1\linewidth]{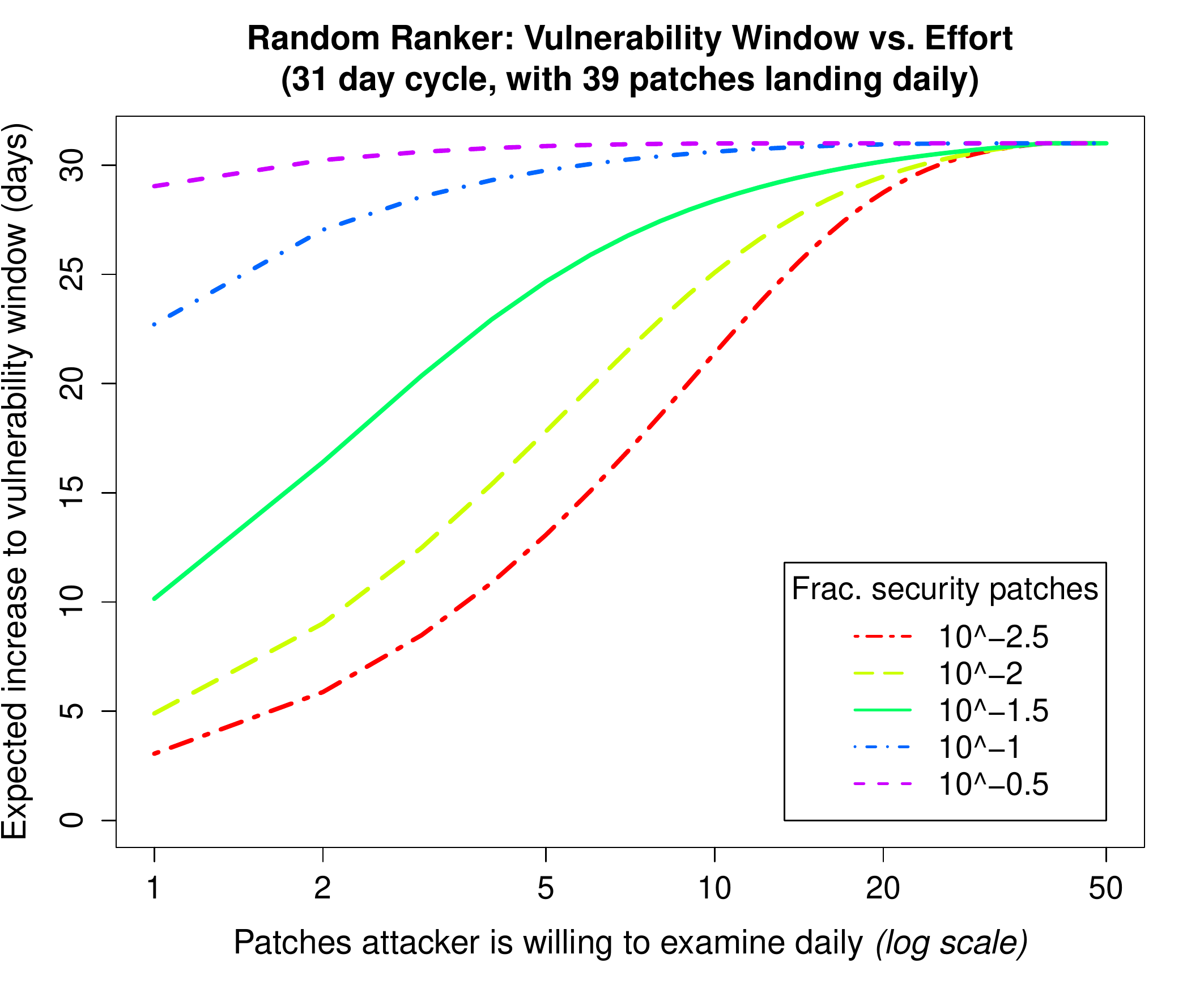}
\caption{The random ranker's expected vulnerability window increase vs. daily budget, for a 31 day cycle with 39 patches daily (the Firefox 3 averages). Benefits shown for security patch fractions of Figure~\ref{fig:random-const-frac-eff}.}
\label{fig:random-const-frac-wind}
\end{minipage}
\end{center}
\end{figure*}

Although these features are harder to obfuscate than the free-form description
field, we do not claim that these features cannot be obfuscated. Instead, we
claim that there are a large number of small information leaks that can be
combined to detect security patches. Of course, this set of features is far
from exhaustive and serves only as a lower bound on the attacker's abilities.

\paragraph{Learning Algorithm.}
For our detection algorithm, we use the popular \LS\ library for support
vector machine~(SVM) learning~\cite{libsvm}. Although we could improve
our metrics by tuning the learning algorithm, we choose to use the default
configuration to strengthen our conclusions---extracting basic features (as
detailed above) and running \LS\ in its default configuration requires only
basic knowledge of Python and almost no expertise in machine learning.

\term{Support vector machines} perform supervised binary classification by
learning a maximum-margin hyperplane in a high-dimensional feature
space~\cite{BurgesTutorial,SVMbook,Kernelbook}. Many feature mappings are
possible, and the default \LS\ configuration uses the feature mapping induced
by the RBF kernel, which takes a parameter $\gamma$ that
controls kernel width. The SVM takes another parameter $C$, which controls
regularization. An attacker need not know how to set these parameters because
\LS\ chooses the parameters that optimize 5-fold cross-validation estimates
over a grid of $(C,\gamma)$ pairs. The optimizing pair is then used to train
the final model. We enable a feature of \LS\ that learns posterior
probability estimates $\Pr{\mbox{\emph{patch fixes a vulnerability}}\mid
\mbox{\emph{patch}}}$ rather than security/non-security class
predictions~\cite{LibSVM-probabilities}. We refer to these posterior
probabilities as \term{probabilities} or \term{scores}.

We present nominal features (author, top-level directory, and file type) to
the SVM as binary vectors. For example, the $i\nth{th}$ author out of $N$
developers in the Firefox project is represented as $N-1$ zeros and
a single $1$ in the $i\nth{th}$ position.

After training an SVM on patches labeled as security or non-security, we can
use the SVM to rank a set of previously unseen patches by ordering the
patches in decreasing order of score. If the SVM is given sufficient training
data, we expect the higher-ranked patches to be more likely to fix
vulnerabilities. As we show in Section~\ref{sec:results}, even though the SVM
scores are unsuitable for classification, they are an effective means for
ranking patches.

\ifthenelse{\isundefined{\fullpaper}}{}{
Note that detecting patches that repair vulnerabilities can be cast as
learning problems other than scalar-valued supervised classification. For
example, we could take a more direct approach via ranking or ordinal
regression (although these again do not directly optimize our primary interest: having
\emph{one} security patch ranked high). However, we use an SVM because it balances statistical performance
for learning highly non-linear decision rules and availability of
off-the-shelf software appropriate for data mining novices.
} 

\paragraph{Online Learning.}
To limit the detector to information available to real attackers, we perform
the following simulation using the dates collected in our data set. For each
day, starting on the day between major releases of Firefox, we perform the following steps:
\begin{enumerate}
\item We train a fresh SVM on all the patches landed in the repository between
the day Firefox~3 was released and the most recent security update before the
current day, labeling each patch according to the publicly known
vulnerabilities list~\cite{known-vulnerabilities}.\footnote{Note that not all security patches are disclosed as fixing vulnerabilities by the following release. Such patches are necessarily (mis)labeled as non-security, and trained on as such. Once the true patch is disclosed, we re-label and re-train. The net effect of delayed disclosure is a slight degradation to the SML-assisted ranker's performance.}
\item We then use the trained SVM to rank all the patches landed since the
most recent security update.
\end{enumerate}
After running the complete online simulation, we observe the highest
ranking received by a real vulnerability fix on each day. This ranking 
corresponds to the SVM-assisted attacker effort for that day.

\subsection{The Random Ranker}

To model an attacker searching for security patches under 
complete metadata obfuscation, we consider a \term{random ranker} who
examines available patches in a random order. By comparing this 
ranker with the SVM, we may gain insight into the amount of information leaked
by seemingly innocuous patch metadata.

We model the random ranker as  selecting patches one-at-a-time,
uniformly-at-random without replacement from the pool of patches available in
\MC. The attacker's effort is the random number $X$ of patches 
examined up to and including the first patch drawn that fixes a
vulnerability. We summarize the cost of using unassisted random ranking via
the \term{expected attacker effort}, which we derive in
\ifthenelse{\isundefined{\fullpaper}}{the full version of this paper}{Appendix~\ref{app:random}}
to be
\begin{eqnarray}
\Exp{X} &=& \begin{pmatrix}n\\ n_s\end{pmatrix}^{-1}\  \sum_{x=1}^{n-n_s+1} x \begin{pmatrix}n-x\\ n_s - 1\end{pmatrix} \label{eq:random-exp} \enspace,
\end{eqnarray}
where $n>0$ is the total number of available patches,
and $1\leq n_s < n$ is the number of these that fix vulnerabilities.

\ifthenelse{\isundefined{\fullpaper}}{}{
The probability mass and expectation of $X$ are explored in
Figures~\ref{fig:random-prob} and~\ref{fig:random-exp}. For $n_s=1$ the
distribution of effort is uniform; and as the number of security patches
increases under a constant pool size, mass quickly concentrates on lower
effort (note that in each figure attacker effort is depicted on a log scale). 
Similarly the significant effect of varying $n_s$ on the expected effort can be
seen in Figure~\ref{fig:random-exp}.
} 

The expected increase to the window of vulnerability is also derived in
\ifthenelse{\isundefined{\fullpaper}}{this paper's full version}{Appendix~\ref{app:random}}.
Both expected cost and benefit metrics can be efficiently
computed exactly for the random ranker, given patch counts $n$ and $n_s$.

Figures~\ref{fig:random-const-frac-eff} and~\ref{fig:random-const-frac-wind}
show the random ranker's expected cost and benefit, for a typical Firefox 3
inter-point release cycle with constant fractions of security patches; in both cases effort is shown on a log scale.
The first figure shows that attacker effort \emph{increases} under a
growing pool of patches, with constant fraction being security-related. For a typical cycle (of length 31 days), typical patch landing rate
(of 39 patches daily) and fixed fraction of security-related landed patches, Figure~\ref{fig:random-const-frac-wind} shows the expected window increase as
a function of daily budget. Again we see a great difference over
increasing proportions of security patches, and the
effect of proportion on the dependence of benefit on budget. 
Since the average fraction of security-related patches for Firefox 3 is
0.0085, the curves at $10^{-2}$ should approximate the
performance of the random ranker for Firefox 3 as is verified in
Section~\ref{sec:results}. By including curves at atypical 
security patch rates (from the perspective of Firefox) we offer a preview of the
cost and benefit achieved by the random ranker as applied to other open-source
projects.

\section{Results} \label{sec:results}

We now present results of searching for Firefox security patches. We first
explore searching using the patch description and \BZ, we then explore the discriminative power of the individual features, and finally we evaluate and compare SVM-assisted search and random search.

\begin{figure}[t]
\begin{center}
\begin{minipage}[t]{1\columnwidth}
\centering
\includegraphics[width=1\textwidth]{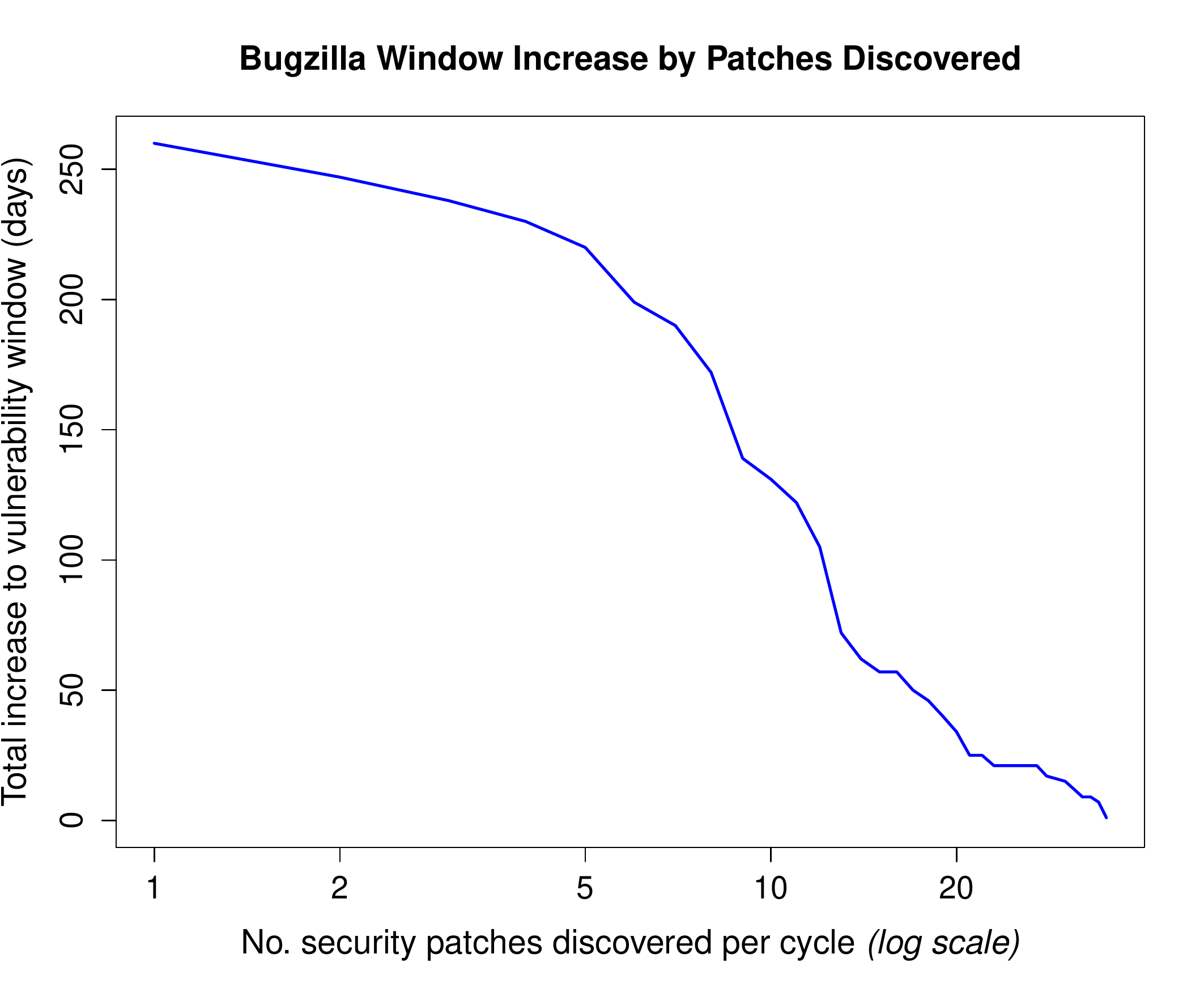}
\caption{The total increases to the window of vulnerability when searching for multiple security patches per inter-release period in Firefox 3, by linking patch description to bug history.}
\label{fig:bug-savings}
\end{minipage}
\end{center}
\end{figure}

\subsection{Exploiting Patch Description and Bug History}

We simulated the search for security patches described in
Section~\ref{sec:bugalgo}, that parses patch description fields for bug
numbers, queries \BZ, and then uses evidence of core-security add/drop
events or access restriction of the bug report pages to determine 
which patches fix vulnerabilities. The attacker effort due to search is essentially
naught---only a handful of  HTTP requests are submitted per landed
patch, and no source code need be analyzed. When executed daily over 300 days of Firefox 3 development, a security patch is found on all but
40 days yielding a total window of vulnerability increase (summed over
all inter-release periods) of a staggering 260 days. 
We obtained similar results on Firefox 3.5 over a period
of 232 days with an aggregate window of vulnerability increase of 219 days.
As an attacker may
wish to find more than one security patch at a time, we re-ran the attack
terminating the search in each inter-release cycle only when the required number of security
patches were discovered. Figure~\ref{fig:bug-savings} displays the
resulting increases to the vulnerability windows as a function of number of security patches desired. The window increase drops slightly to 247 and
238 days total, when the required number of security patches grows to
2 and 3 respectively.

\emph{By exploiting patch descriptions and the nature of security patches
leaked by the Firefox bug tracker, an attacker searching for security patches
increases the total window of vulnerability for Firefox 3 and 3.5 by factors of
9.4 and 10.1 over baseline, respectively.}

\subsection{Metadata Feature Analysis} \label{subsec:results-features}

\begin{figure}[t]
\begin{center}
\begin{minipage}[t]{1\columnwidth}
\centering
\includegraphics[width=1\linewidth]{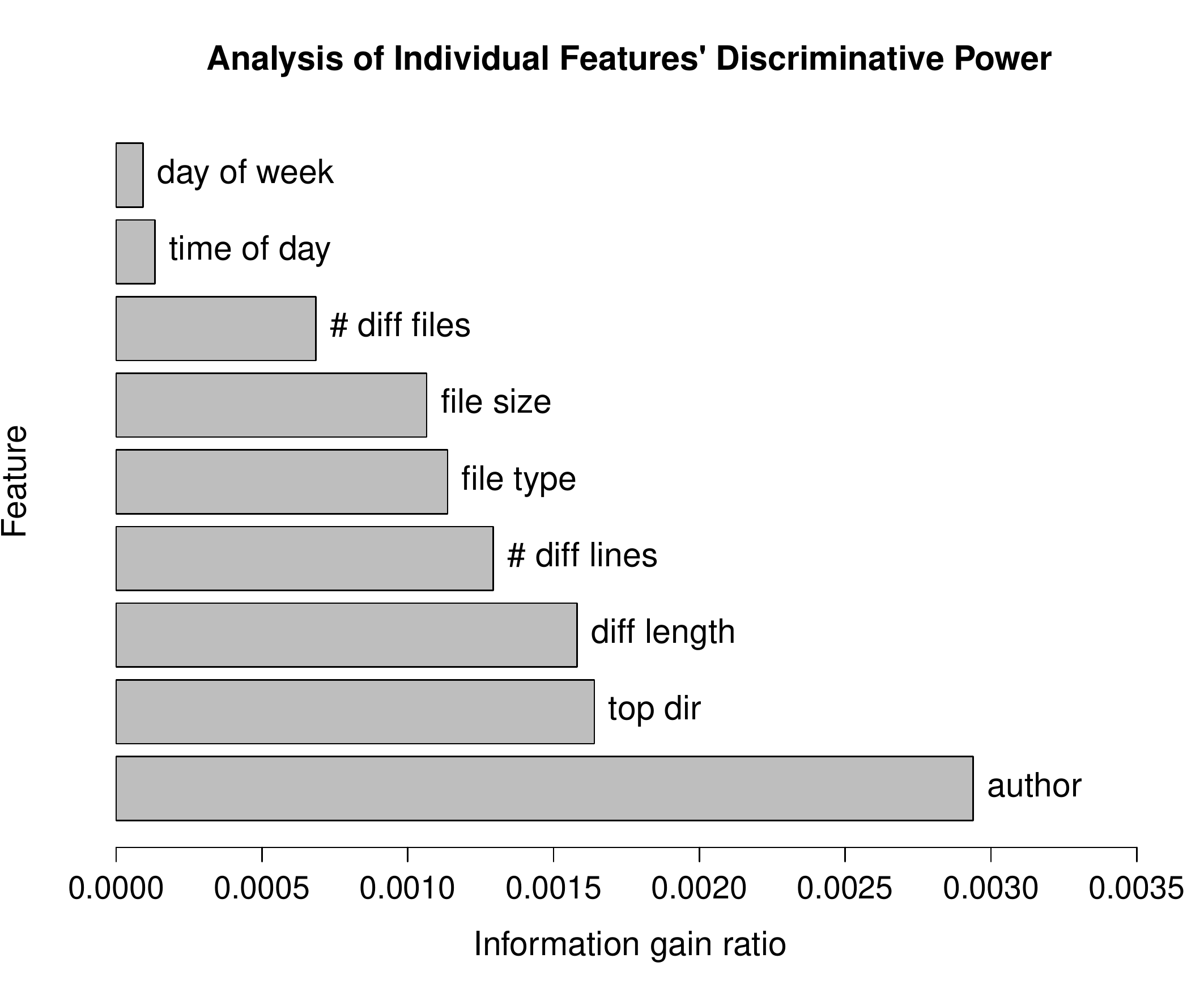}
\caption{The features ordered by decreasing ability to discriminate between security and non-security patches, as represented by the information gain ratio.}
\label{fig:info-gain-ratio}
\end{minipage}
\end{center}
\end{figure}

\begin{figure*}[t]
\begin{center}
\begin{minipage}[t]{.32\textwidth}
\centering
\includegraphics[width=1\textwidth]{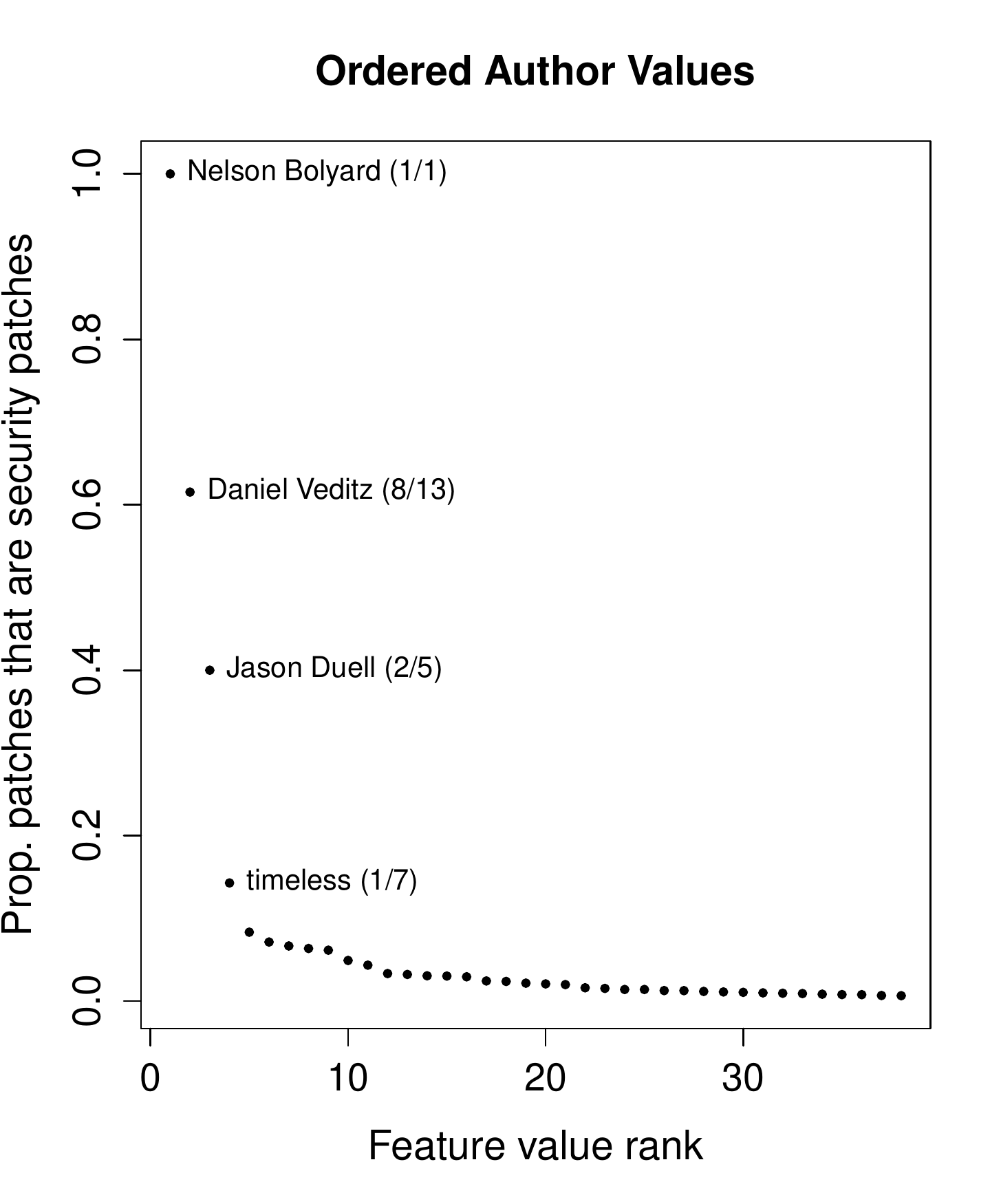}
\caption{The authors who committed security patches, ordered by proportion of patches that are security patches The top four authors are identified with their security and total patches along-side.}
\label{fig:featanal-author}
\end{minipage}\hfill
\begin{minipage}[t]{.32\textwidth}
\centering
\includegraphics[width=1\textwidth]{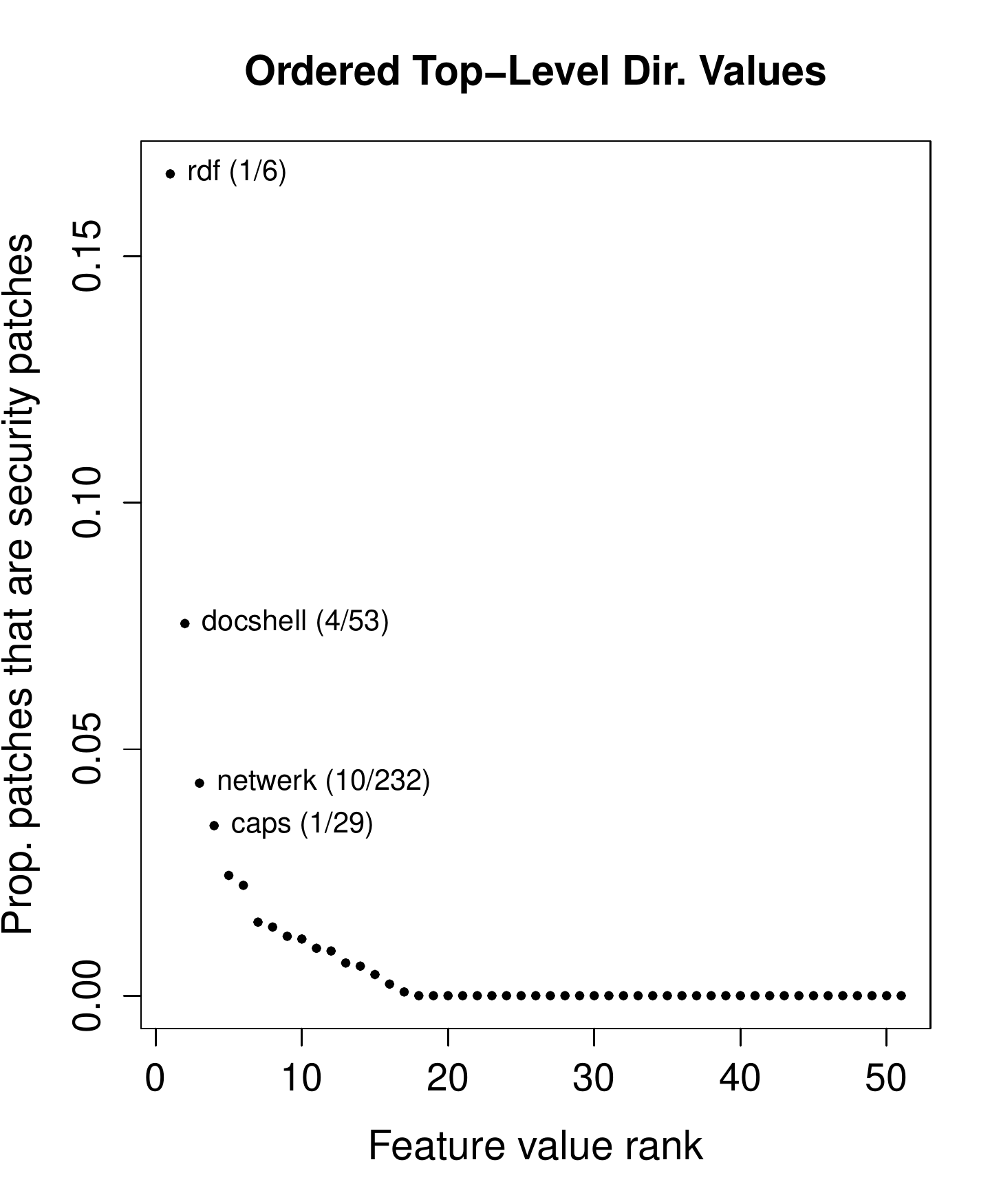}
\caption{The top-level directories ordered by the proportion of patches that are security patches The top four directories are identified with their security and total patches along-side.}
\label{fig:featanal-topdir}
\end{minipage}\hfill
\begin{minipage}[t]{.32\textwidth}
\centering
\includegraphics[width=1\textwidth]{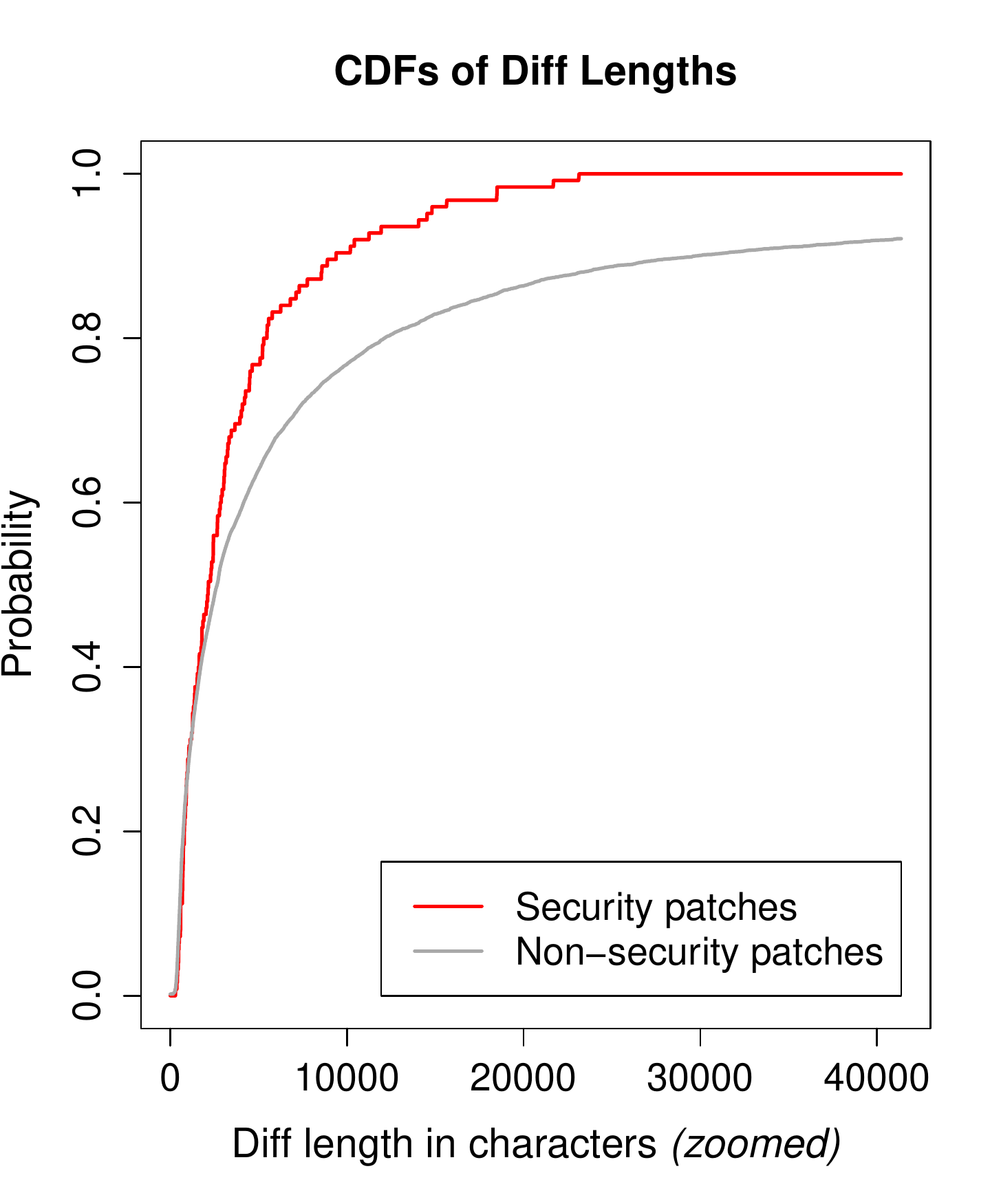}
\caption{The CDFs of the security and non-security diff lengths. The figure is
``zoomed'' in on the left, with the top $1,\!000$ largest lengths not shown.}
\label{fig:featanal-difflen}
\end{minipage}
\end{center}
\end{figure*}

Prior to evaluating the SVM-assisted ranker, we analyzed the
ability of individual features to discriminate between security and
non-security patches. We adopt the information theoretic
\term{information gain ratio}, which reflects the decrease in entropy
of the training set class labels when split by each
individual feature\ifthenelse{\isundefined{\fullpaper}}{}{
(\cf\ Appendix~\ref{sec:infogain} for details on information gain)}.
The results are presented in
Figures~\ref{fig:info-gain-ratio}--\ref{fig:featanal-difflen}.

The individual
features' discriminative abilities are recorded in
Figure~\ref{fig:info-gain-ratio}. For the nominal features---author, top-level
directory, file type, and day of week---we compute the information gain ratios
directly, whereas for the remaining continuous features 
we use the gain ratio given by choosing the best threshold value.
The author feature has
the most discriminative power, providing a gain ratio $1.8$ times
larger than the next most informative feature. The next two most
discriminative features are top-level directory and diff length, with the remaining features contributing less information.%
\ifthenelse{\isundefined{\fullpaper}}{
\emph{Some features provide discriminative power for separating security
  patches from non-security patches, with author, top-level directory,
  and diff length among the most discriminative.}
}{} 

Furthermore, observe that%
\ifthenelse{\isundefined{\fullpaper}}{
\emph{each individual feature alone provides insignificant discriminative power}}{each individual feature alone provides only insignificant discriminative power}, 
since the maximum information gain ratio is a tiny $3\times 10^{-3}$.
\ifthenelse{\isundefined{\fullpaper}}{}{
To add credence to these numbers, we note also that the unnormalized
information gains (\cf\ Appendix~\ref{sec:infogain}) have a similar
ordering with the author feature coming out on top with an information
gain of $2\times 10^{-2}$, which corresponds to a small change in entropy.
To summarize, we offer the following remark.

\begin{remark}
  Some features provide discriminative power for separating security
  patches from non-security patches, with author, top-level directory,
  and diff length among the most discriminative. However,
  individually, no feature provides significant discriminative power
  for separating security patches from non-security patches.
\end{remark}
}

To give intuition why some features provide discriminating power for
security vs. non-security patches, we analyze in more detail the three most
discriminative features: the author, top-level directory, and diff
length. 

For the author and top-level directory features, we analyze their
influential feature values. For each occurring feature value, we
compute its \textit{proportion value}: the proportion of 
patches with that feature value that are security sensitive. Figures~\ref{fig:featanal-author}
and~\ref{fig:featanal-topdir} depict the influential feature values for 
authors and top-level directory by ranking the feature values
by their proportion values. During the life-cycle of Firefox~3, a
total of $516$ authors contributed patches out of which $38$
contributed at least one security patch; only these $38$ authors are
shown in Figure~\ref{fig:featanal-author}. Notice that the top four
authors (labeled) are all members of the Mozilla Security
Group~\cite{mozilla-security-group}.

To explore the third most individually discriminative feature, the continuous
diff length, we plot the feature CDFs for security and non-security patches in
Figure~\ref{fig:featanal-difflen}. Since the diff length distribution is 
extremely tail heavy (producing CDFs resembling step-functions) Figure~\ref{fig:featanal-difflen} zooms in on the left portion of
the CDF by plotting the curves for the first $6,500$ (out of $7,500$) unique
diff lengths. From the relative positions of the CDFs, we observe that
security patches have shorter diff lengths than non-security patches. This
matches our expectations that patches that add features to Firefox require
larger diffs.

\begin{figure*}[t]
\begin{center}
\begin{minipage}[t]{1\textwidth}
\centering
\includegraphics[width=1\textwidth]{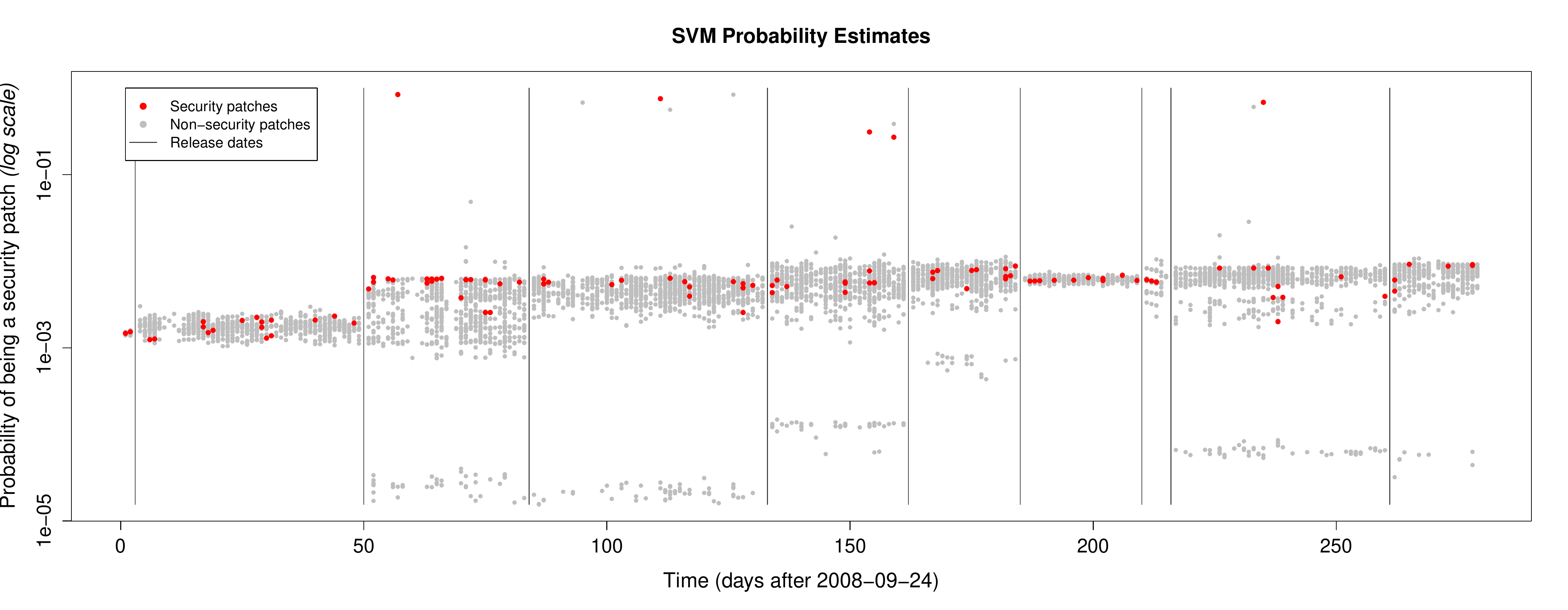}
\caption{The time series of SVM probability estimates, with security patches and
non-security patches delineated by color.}
\label{fig:svm-scores}
\end{minipage}
\end{center}
\end{figure*}

\begin{figure*}[t]
\begin{center}
\begin{minipage}[t]{1\textwidth}
\centering
\includegraphics[width=1\textwidth]{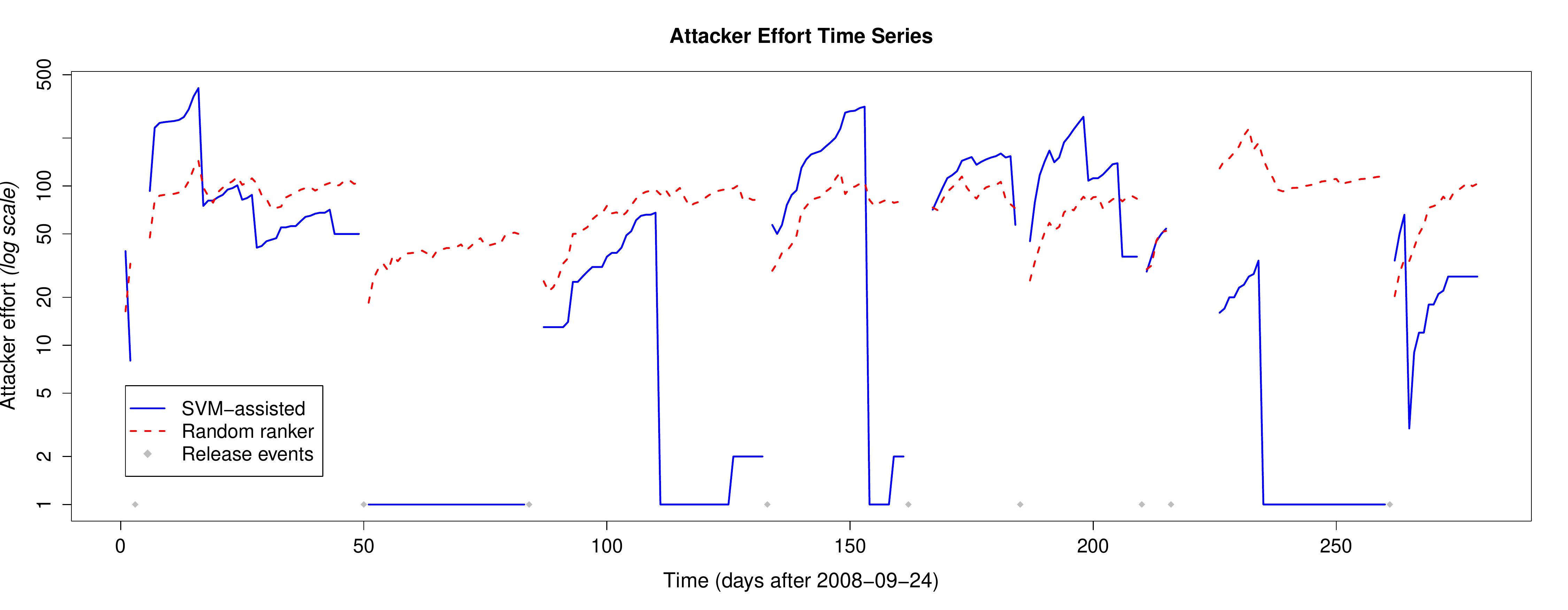}
\caption{The attacker effort of the SVM, and the expected attacker effort of the random
ranker, as a function of time.}
\label{fig:all-series-o1}
\end{minipage}
\end{center}
\end{figure*}

Our feature analysis shows that no individual feature effectively predicts
whether a patch is security-related: motivated
attackers should look to more sophisticated statistical techniques to reduce
the cost of finding security patches; and suggests that obfuscating individual features will not plug information leaks in the open-source
life-cycle. Moreover certain features cannot be effectively obfuscated. For
example, the Mozilla Committer's Agreement would be violated if
developer names were redacted from \MC.

\subsection{Classifier Performance} \label{subsec:results-classifier}

Figure~\ref{fig:svm-scores} depicts the time series of scores assigned
to each patch by the SVM (the horizontal axis shows the day when each patch is
landed in the repository starting at 2008-09-24; at which point security patches
were first announced). Note that as mentioned in
Section~\ref{sec:svmalgo}, we use an online learning approach, so
the score assigned to a patch is only computed using the SVM trained
with labeled patches seen up to the most recent security update prior
to the patch landing in the repository (and including any out-of-release
delayed disclosures occurring before the patch is landed). Notice that the scores
for security and non-security patches in the first $50$ days are
quite similar. Over time, the SVM learns to assign high scores to a
handful of vulnerability fixes (and a few non-security patches) and low
scores to a handful of non-security patches. However, many patches are
assigned very similar scores of around $0.01$ irrespective of whether
they fix vulnerabilities.

Viewed as a binary classifier, the SVM performs poorly because
there is no sharp threshold that divides security patches from non-security
patches. However, when viewed in terms of the attacker's utility, the SVM
might still be useful in reducing effort because the \emph{relative} rankings
of the vulnerability fixes are generally higher than most non-security
patches.  

\subsection{Cost-Benefit Analysis of SVM and Random Rankers}
\label{subsec:results-detector}

\paragraph{Attacker Effort.}
Figure~\ref{fig:all-series-o1} shows the time series of the effort the
attacker expends to find a vulnerability (as measured by the number of
patches the attacker examines), as described in
Section~\ref{sec:attacker-effort}.  The attacker effort measured for a
given day is computed to reflect the following estimate.  Imagine, for
example, an attacker who ``wakes up'' on a given day, trains an SVM on
publicly available information (including all labeled patches before the
current day), and then starts
looking for security patches among all the (unlabeled) patches landed in the
repository since the most recent security update, in rank order
provided by the SVM. Then the attacker effort measured for a given day is
the number of patches that the attacker has to examine before finding
a security patch using the rank order provided by the SVM.

Each continuous segment in the graph corresponds to one of the $12$ security
updates during
our study. For a period of time after each release, there are
no security patches in \MC, which is represented on the graph as a
gap between the segments. For the first $50$ days of the experiment both the
random ranker and the SVM-assisted attacker expend
relatively large amounts of effort to find security patches. This poor initial
performance of the SVM, also observed in Figure~\ref{fig:svm-scores}, is due
to insufficient training. 
The SVM, like any statistical estimator, requires enough data with which to
generalize.
Given the SVM's ``warm-up'' effect during the first $50$
days, all non-time-series figures in the sequel are shown using data after 2008-11-13 only.

\begin{figure}[t]
\begin{center}
\begin{minipage}[t]{1\columnwidth}
\centering
\includegraphics[width=1\textwidth]{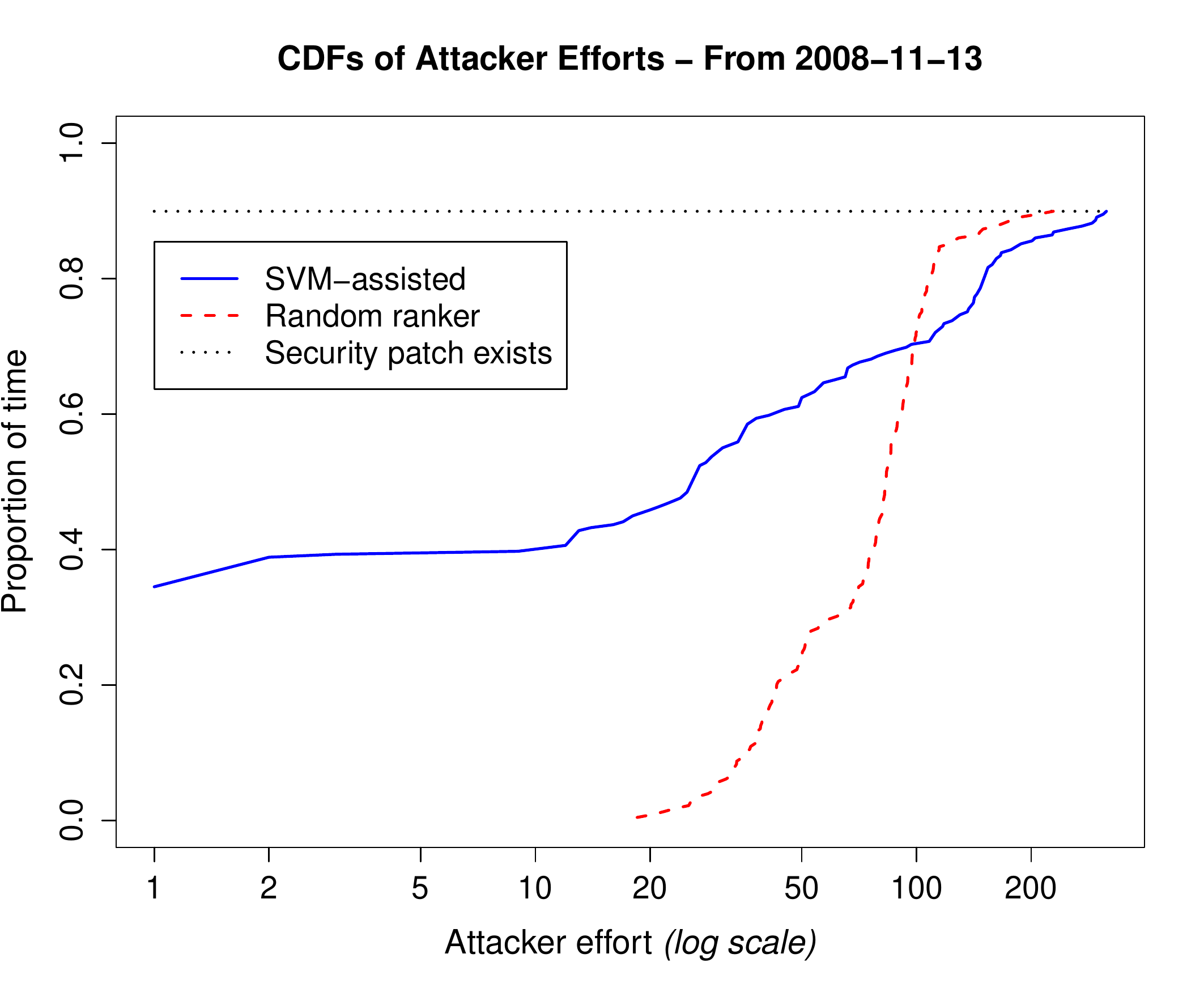}
\caption{The cumulative distribution functions of the attacker effort displayed
in Figure~\ref{fig:all-series-o1}, from 11/13/2008 onwards. CDFs are shown for
both the SVM and the random ranker.}
\label{fig:all-CDF-o1}
\end{minipage}
\end{center}
\end{figure}

\begin{remark}
  During the latter 2/3\nth{rds} of the year (the 8 month period starting 50 days after 2008-09-24) the SVM-assisted attacker, now with
  enough training data, regularly expends significantly less
  effort than an attacker who examines patches in a random order.
\end{remark}

The general cyclic trends of the SVM-assisted and random rankers are also
noteworthy. In most inter-update periods, the random ranker enjoys a
relatively low attacker effort (though higher than the SVM's) which quickly
increases.
\ifthenelse{\isundefined{\fullpaper}}{The reason for this behavior can be seen in Figure~\ref{fig:random-const-frac-eff},}{The reason for this behavior can
be understood by plotting the expected effort for the random ranker with
respect to the number of security patches for various total patch pool sizes
as shown in Figure~\ref{fig:random-exp}. Immediately after the landing of
a first post-update security patch, the pool of available patches gets
swamped by non-security patches (\cf\ Figure~\ref{fig:thundering-herd}), corresponding to increasing $n$ in Figure~\ref{fig:random-exp} and greatly
increasing the expectation. Further landings of security patches are few and
far between (by virtue of the rarity of such patches), and so moving across
the figure with increasing $n_s$ is rare. As the periods progress,
non-security patches continue to swamp security patches. This trend for the random ranker's expected effort is more directly
seen in Figure~\ref{fig:random-const-frac-eff},} 
which plots expected effort over
a prototypical cycle of Firefox 3. Over the single 31 day cycle, 39 patches
land daily of which a constant proportion are security patches. The curve for
$10^{-2}$ most closely represents Firefox 3 where the security patch rate is
$0.0085$ of the total patch rate. The trend observed empirically in
Figure~\ref{fig:all-series-o1} matches both the overall shape and location of
the predicted trend.
\ifthenelse{\isundefined{\fullpaper}}{\emph{Immediately after the landing of a first post-update security patch, the pool of available patches gets swamped by non-security patches. This situation continues, resulting in trends of increasing random ranker expected effort.}}{} 

\begin{figure}[t]
\begin{center}
\begin{minipage}[t]{1\columnwidth}
\centering
\includegraphics[width=1\textwidth]{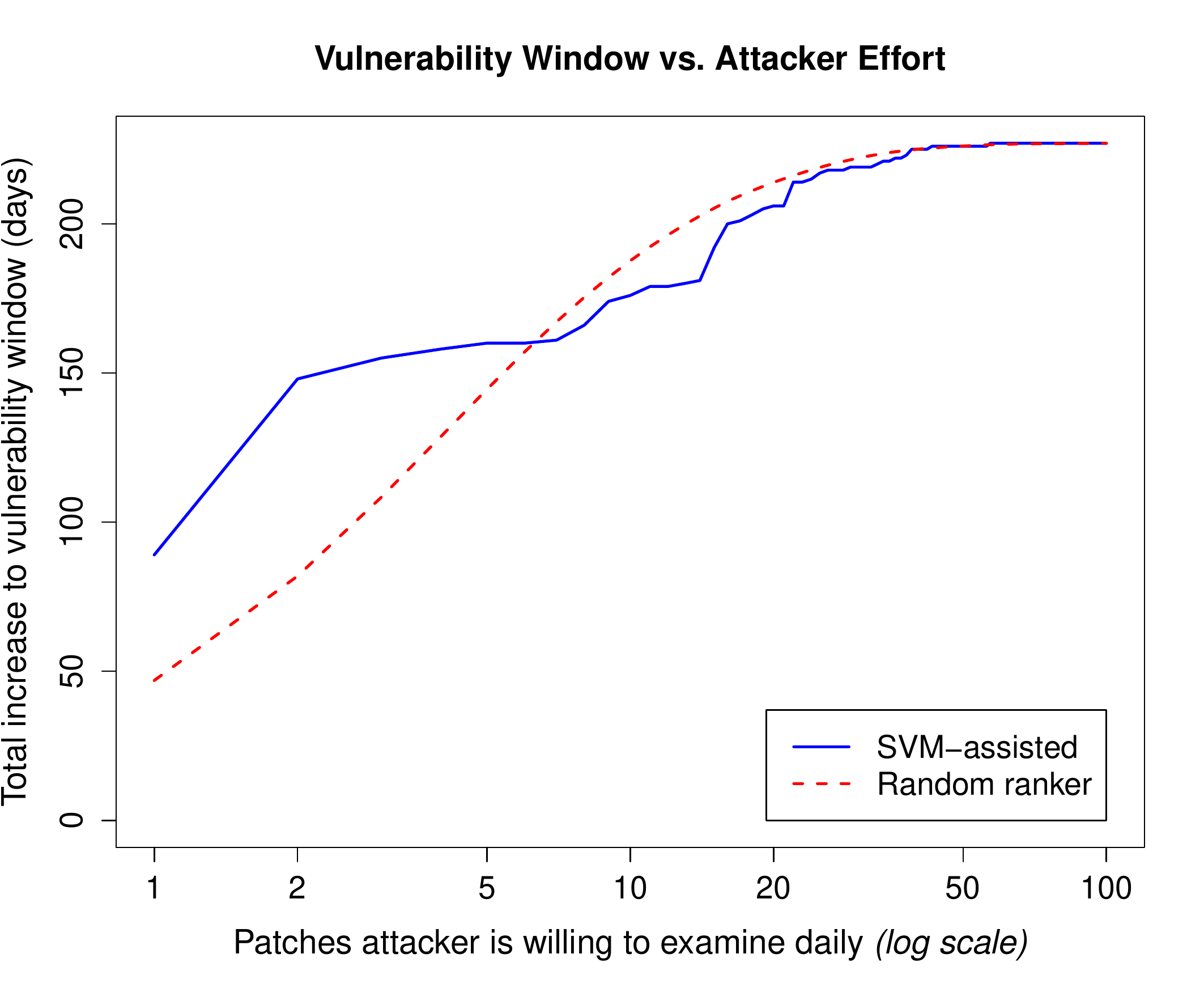}
\caption{The total increase to the vulnerability window size throughout the
year, for a given level of daily attacker effort with or without SVM assistance.
Results trimmed to 11/13/2008 onwards.}
\label{fig:vul-wind-increase}
\end{minipage}
\end{center}
\end{figure}

At times early on in the inter-release periods, the SVM-assisted
attacker experiences the same upward trending effort, but eventually the
developers land a security patch that resembles the training data. Given just
one ``easy'' fix, the effort required of the SVM-assisted attacker plummets. In 
two cycles (and partially in two others) the SVM-assisted attacker must 
expend more effort than the random ranker. This is due to a combination of 
factors including the small rates of landing security patches which means that
unreleased security patches may not resemble training data.

\paragraph{Proportion of Days of Successful Vulnerability Discovery.}
Figure~\ref{fig:all-CDF-o1} depicts the cumulative distribution function (CDF)
of attacker effort, showing how often the SVM-assisted and random rankers can
find a security patch as a function of effort. Note that the CDFs asymptote to
$0.90$ rather than $1.0$ because \MC\ did not contain any security patches
during $10\%$ of the $8$ month period.

\begin{remark}
The SVM-assisted attacker discovers a security patch with the first examined
patch for $34\%$ of the $8$ month period. 
\end{remark}

If the unassisted attacker expends the minimum effort of $18.5$, it
can only find security patches for less than  $0.5\%$ of the $8$ month period. By
contrast, an SVM-assisted attacker who examines $17$ patches will find
a security patch during $44\%$ of the period. In order to find
security patches for $22\%$ of the $8$ month period, the random ranker
must examine on average up to $70.3$ patches. \emph{The SVM-assisted
  attacker achieves significantly greater benefit than an attacker who examines patches in random order, when small to moderate numbers of patches are examined (\ie\ up to 100 patches).}

\begin{figure*}[t]
\begin{center}
\begin{minipage}[t]{1\textwidth}
\centering
\includegraphics[width=1\textwidth]{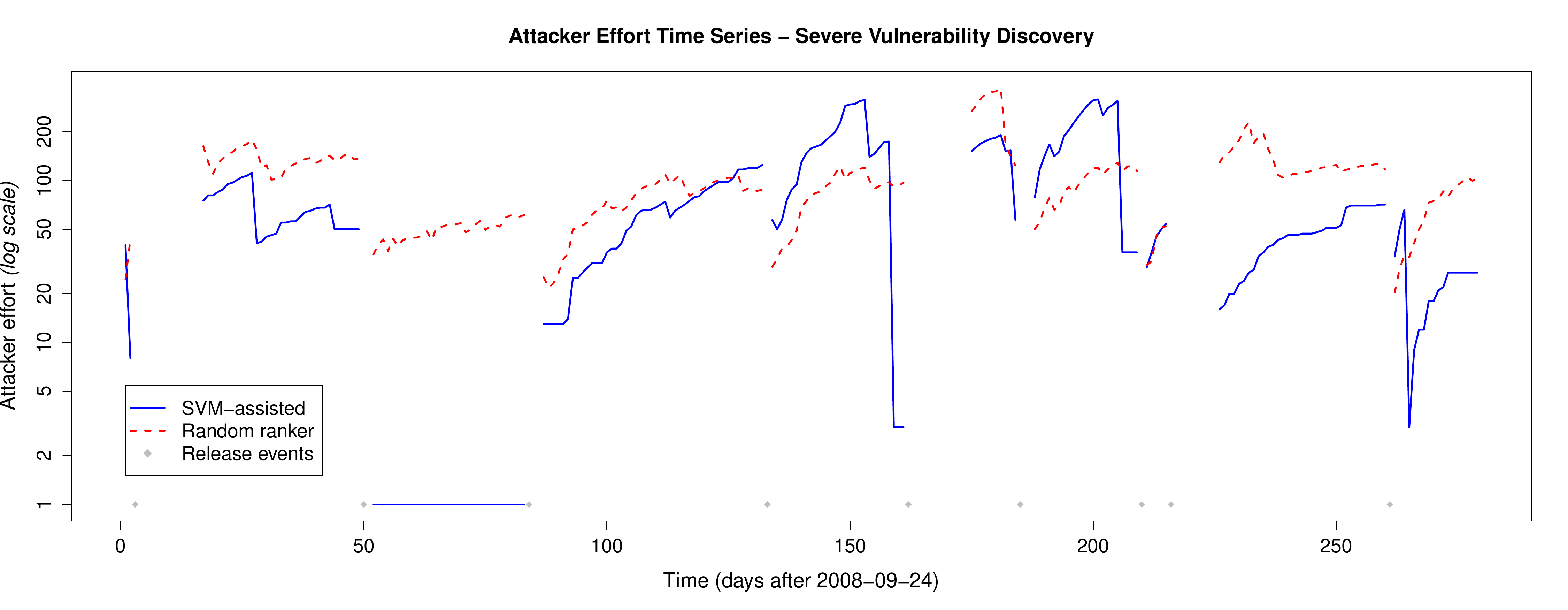}
\caption{The time series of SVM-assisted and random ranker effort for
finding severe (high or critical level) vulnerabilities.}
\label{fig:severe-series}
\end{minipage}
\end{center}
\end{figure*}

\begin{figure*}[t]
\begin{center}
\begin{minipage}[t]{.485\textwidth}
\centering
\includegraphics[width=1\textwidth]{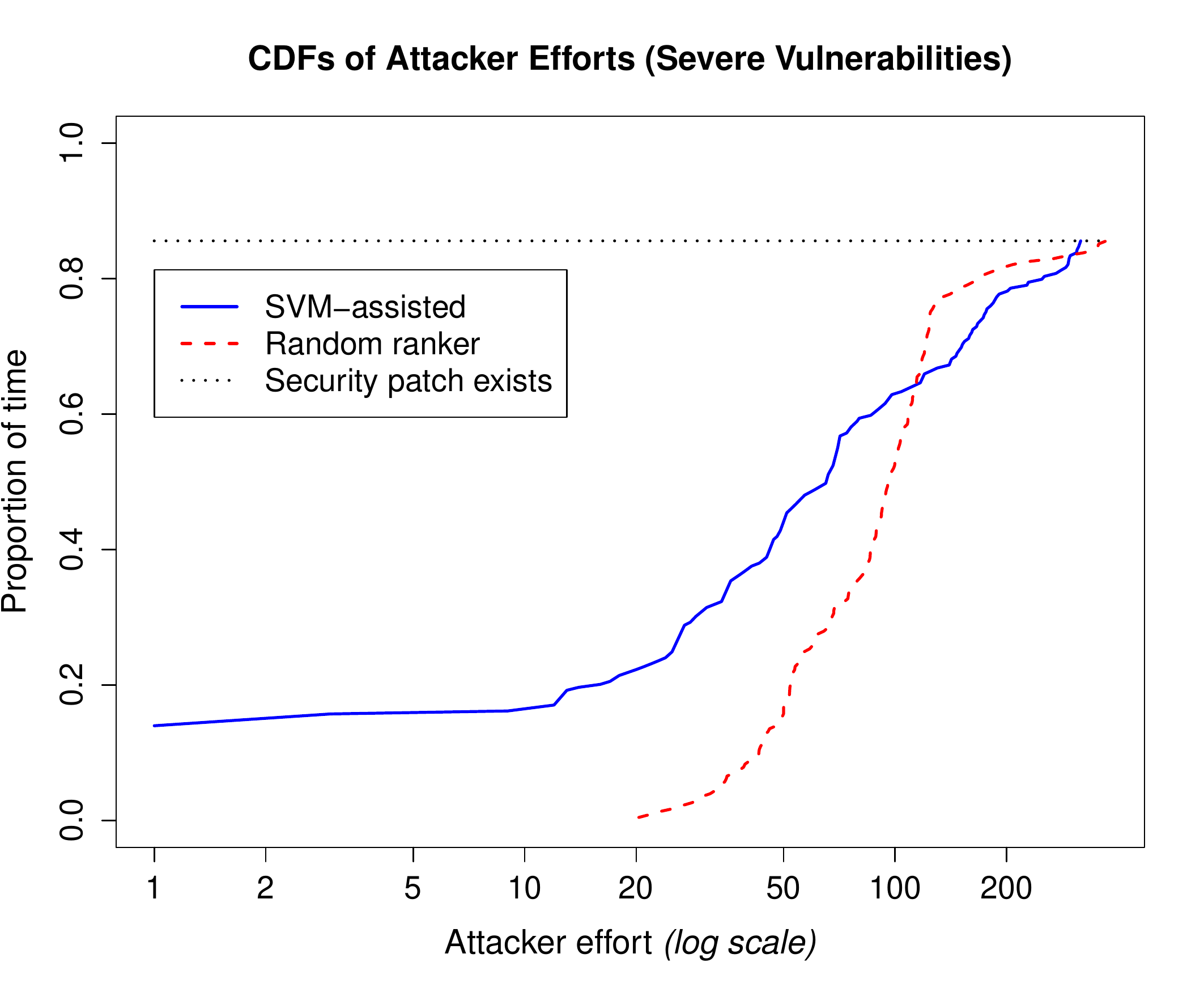}
\caption{The CDFs of the SVM-assisted and random ranker efforts for discovering severe vulnerabilities, from 11/13/2008 onwards.}
\label{fig:severe-cdf}
\end{minipage}\hfill
\begin{minipage}[t]{.485\textwidth}
\centering
\includegraphics[width=1\textwidth]{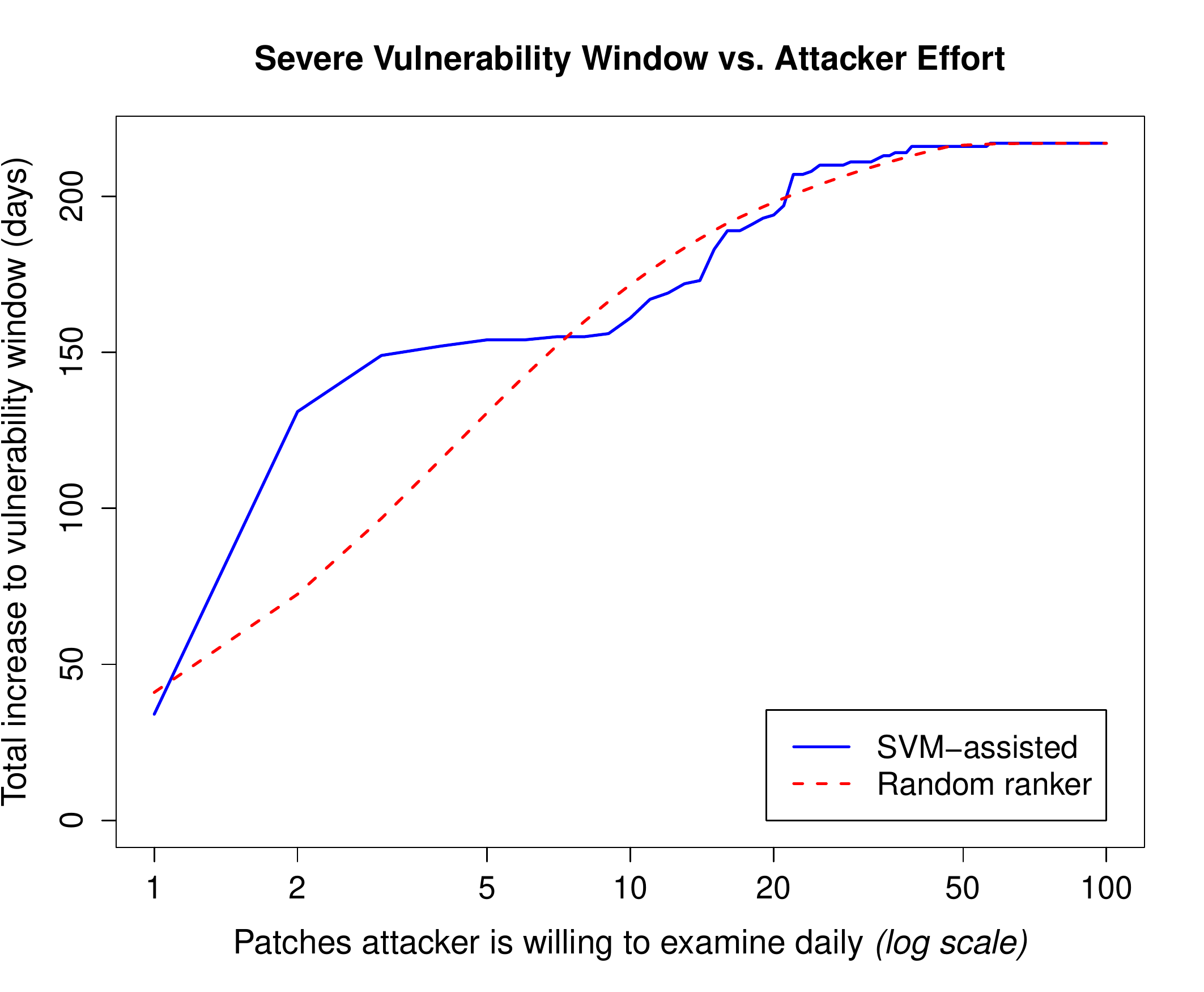}
\caption{Total increase to the vulnerability window for finding severe vulnerabilities given levels of daily attacker effort, from 11/13/2008 onwards.}
\label{fig:severe-vuln-window}
\end{minipage}
\end{center}
\end{figure*}

When examining 100 or more patches, the SVM-assisted and random rankers
find security patches for similar proportions of the 8 month period, with the
random ranker achieving slightly better performance.

\paragraph{Total Increase to the Window of Vulnerability.}
While the CDF of attacker effort measures how hard the attacker
must work in order to find a patch that fixes a vulnerability, 
Figure~\ref{fig:vul-wind-increase} estimates the value of
discovering a vulnerability by measuring the total increase to the
window of vulnerability gained by an attacker who expends a given
amount of effort each day (\cf\ Section~\ref{subsec:vulnerability-window}). Note that this differs from the
previous section by considering an attacker who aggregates work over
multiple days, and who does not re-examine patches from day-to-day.

\begin{remark}
At $1$ or $2$ patches examined daily over the $8$ month period, the
SVM-assisted attacker increases the window of vulnerability by $89$ or $148$
days total, respectively. By contrast the random ranker must examine
$3$ or $7$ patches a day (roughly $3$ times the work) to achieve
the approximate same benefit. At small budgets of $1$ or $2$ patches daily, the
random ranker achieves window increases of $47$ or $82$ days which are just over
half the SVM-assisted attacker's benefits. At higher daily budgets of 7 patches
or more, the two attackers achieve very similar benefits with the random
ranker's being insignificantly greater.
\end{remark}

\begin{figure*}[t]
\begin{center}
\begin{minipage}[t]{1\textwidth}
\centering
\includegraphics[width=1\textwidth]{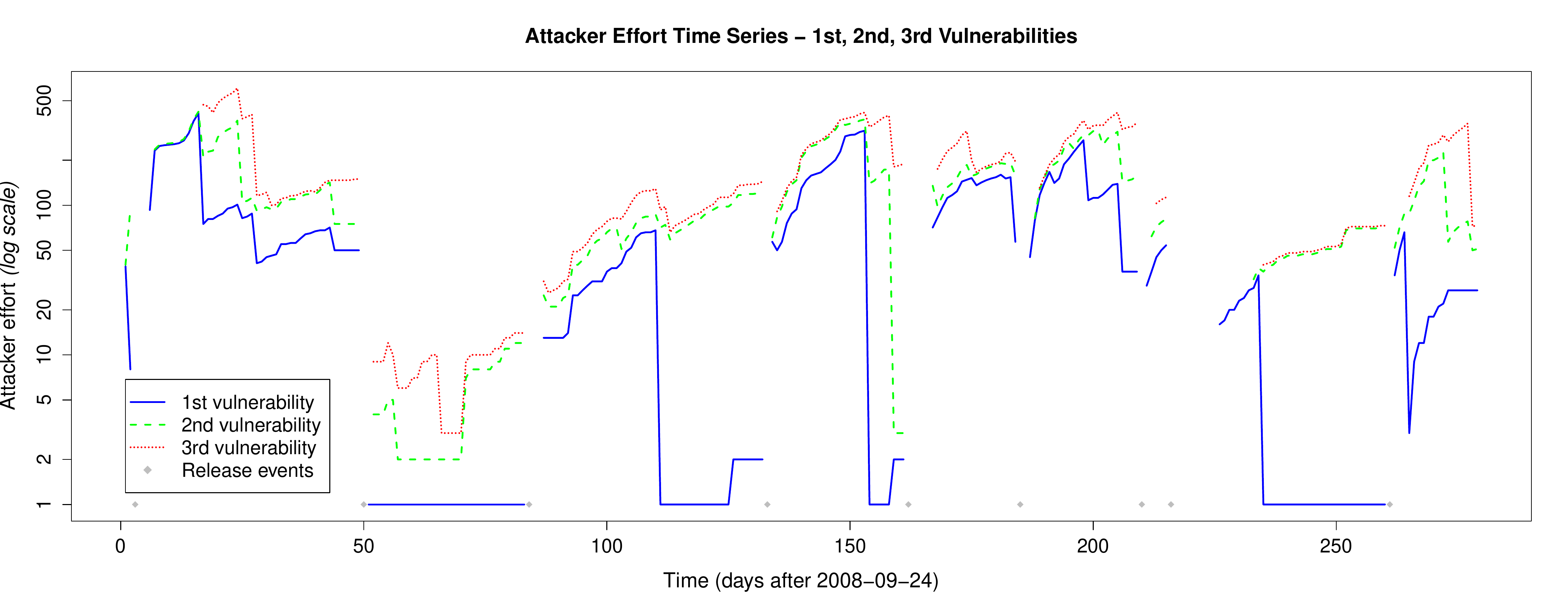}
\caption{The time series of SVM-assisted ranker effort for
finding 1, 2 or 3 vulnerabilities.}
\label{fig:multiple-series}
\end{minipage}
\end{center}
\end{figure*}

Compared to the Firefox 3 base-line vulnerability window size of 3.4 days
(\cf\ Section~\ref{subsec:vulnerability-window}), the increases to window size of
89 and 148 represent multiplicative increases by factors of 3.9 and 6.4
respectively.

\paragraph{In Search of Severe Vulnerabilities.}
Thus far, we have treated all vulnerabilities equally. In reality, attackers
prefer to exploit higher severity vulnerabilities because those
vulnerabilities let the attacker gain greater control over the user's system. To
evaluate how well the attacker fairs at finding severe vulnerabilities---those
judged as either ``high'' or ``critical'' in
impact~\cite{mozilla-severity}---we measure the attacker effort required to
find the first high or critical vulnerability (\ie\ we ignore ``low'' and
``moderate'' vulnerabilities). Note that we did not re-train the SVM on severe
vulnerabilities even though re-training could lead to better results for the
special case of discovering high-severity vulnerabilities.
Figures~\ref{fig:severe-series}--\ref{fig:severe-vuln-window} present our
results for finding severe vulnerability fixes. The attacker effort time
series for the SVM-assisted and random rankers are displayed in
Figure~\ref{fig:severe-series}. Overall, attacker effort curves are similar 
for all vulnerabilities, but shifted upwards away from 1 during several
inter-update periods.

We can interpret the effect of focusing on severe vulnerabilities by examining
the attacker effort CDFs in Figure~\ref{fig:severe-cdf}. Although both
attackers asymptote to the lower proportion of the period containing severe
vulnerability fixes (down from $90\%$ for identifying arbitrary
vulnerabilities to $86\%$), only the random ranker's CDF is otherwise
relatively unchanged. The random ranker's minimum effort has increased from
$18.5$ to $20.3$ patches with a similarly low probability. The SVM-assisted
attacker CDF undergoes a more drastic change. Examining one patch results in a
vulnerability for $14\%$ of the $8$ month period, whereas an effort of
$6$ and $21$ produce vulnerabilities for $20\%$ and $34\%$ of
the $8$ month period, respectively. To achieve these three proportions the
random ranker must examine $48$, $52$, and $76$ patches.

\begin{figure*}[t]
\begin{center}
\begin{minipage}[t]{.485\textwidth}
\centering
\includegraphics[width=1\textwidth]{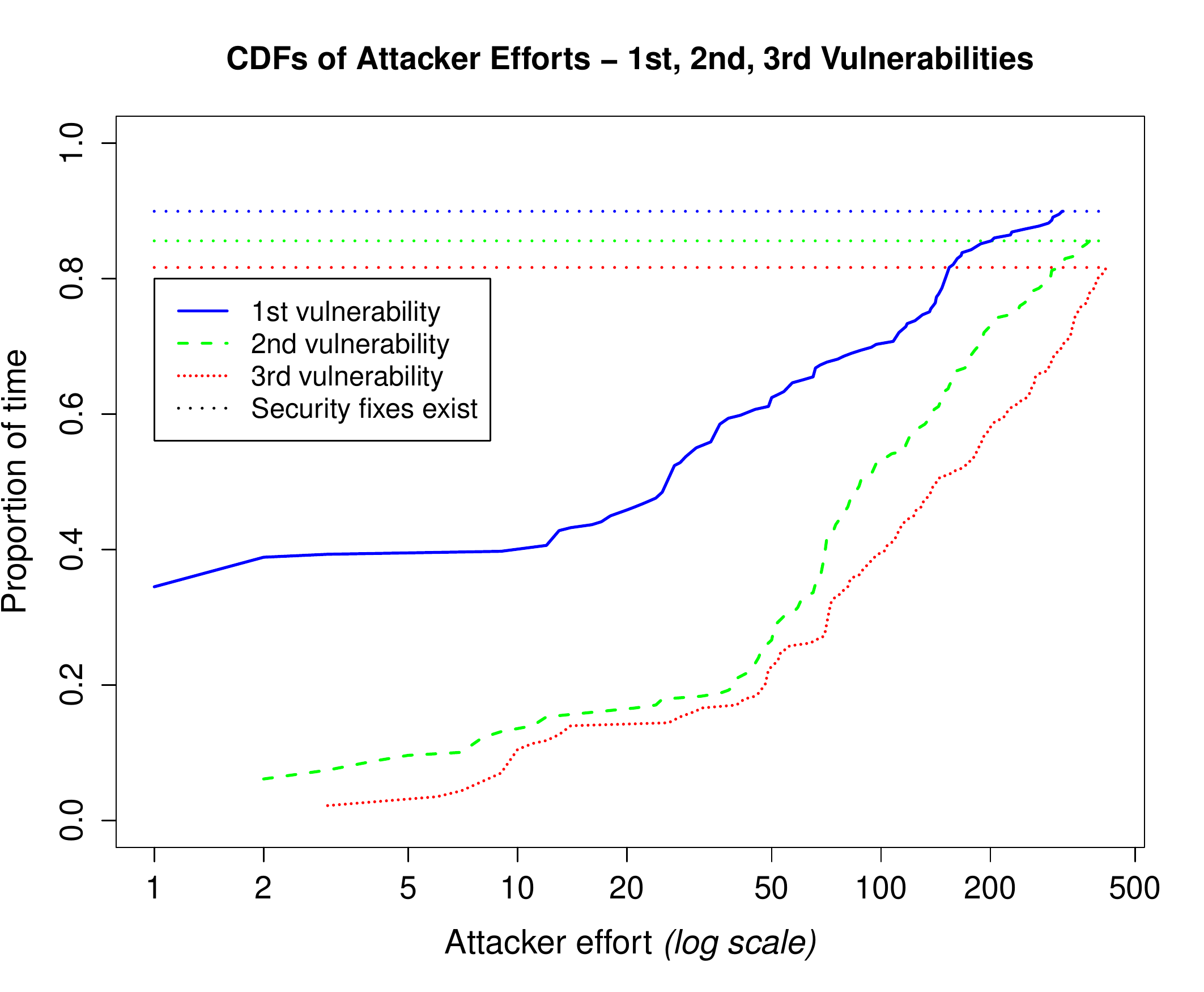}
\caption{The CDFs of the SVM-assisted efforts for discovering 1, 2 or 3 vulnerabilities, from 11/13/2008 onwards.}
\label{fig:multiple-cdf}
\end{minipage}\hfill
\begin{minipage}[t]{.485\textwidth}
\centering
\includegraphics[width=1\textwidth]{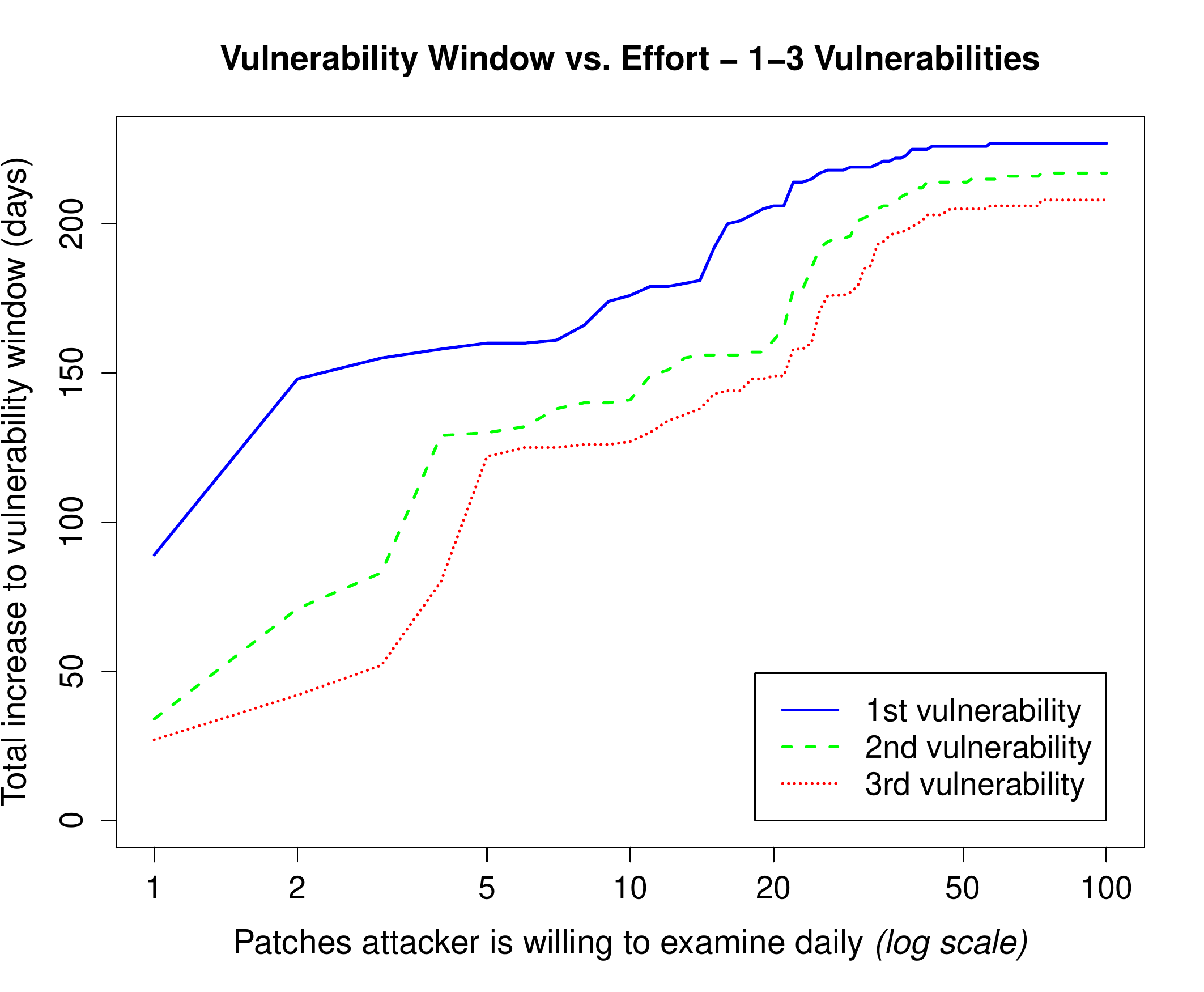}
\caption{Total increase to the vulnerability window for finding 1, 2 or 3 vulnerabilities given levels of daily attacker effort, from 11/13/2008 onwards.}
\label{fig:multiple-vuln-window}
\end{minipage}
\end{center}
\end{figure*}

\begin{remark}
The SVM-assisted attacker is still able to outperform the random ranker in
finding severe vulnerabilities, in particular finding such security fixes
$20\%$ of the time by examining $6$ patches.
\end{remark}

The increases to the severe vulnerability window are shown for the two
attackers in Figure~\ref{fig:severe-vuln-window}. Again, we see a
shift, with the SVM-assisted attacker continuing to outperform the random
ranker on small budgets (except for a budget of 1 patch) or otherwise perform
similarly.
\begin{remark}
By examining $2$ patches daily during the $8$ month period, the
SVM-assisted attacker increases the vulnerability window by $131$ 
days. By contrast the random ranker with budget $2$ achieves an expected window
increase of $72$ days.
\end{remark}

\paragraph{Searching for Multiple Vulnerabilities.}
An attacker searching for security patches might suffer from false
negatives: the attacker might mistakenly take a security patch as a
non-security patch; or an attacker may simply wish to examine
more patches than represented by the attacker effort defined above. To
model this situation, we considered the problem of finding $2$ or $3$
security patches instead of just one.
 
As depicted in
Figures~\ref{fig:multiple-series}--\ref{fig:multiple-vuln-window},
finding $1$, $2$, or $3$ security patches
requires progressively more effort. When computing the increase to the
window of vulnerabilities in Figure~\ref{fig:multiple-vuln-window}, we
assume that the attacker's analysis of the examined patches only turns
up the 1\nth{st}, 2\nth{nd} and 3\nth{rd} security fixes
respectively. To find $2$ or $3$ security patches over $34\%$ of the
$8$ month period, the SVM-assisted attacker must examine $35$ or $36$
patches respectively.

Finally consider approximating the window of vulnerability achieved by
an attacker examining a single patch daily with no false negatives.
Examining $3$ patches a day increases
the total vulnerability window by $83$ days even if the attacker's
analysis produces one false negative each day. Assuming two false
negatives each day, examining $4$ patches daily increases the window by $80$
days total. Similarly increasing the window by $151$ or $148$ days,
approximating the error-free result under a two patch per day budget, requires
examining $12$ or $18$ patches daily when suffering one or two false negatives
respectively.

\begin{figure*}[t]
\begin{center}
\begin{minipage}[t]{1\textwidth}
\centering
\includegraphics[width=1\textwidth]{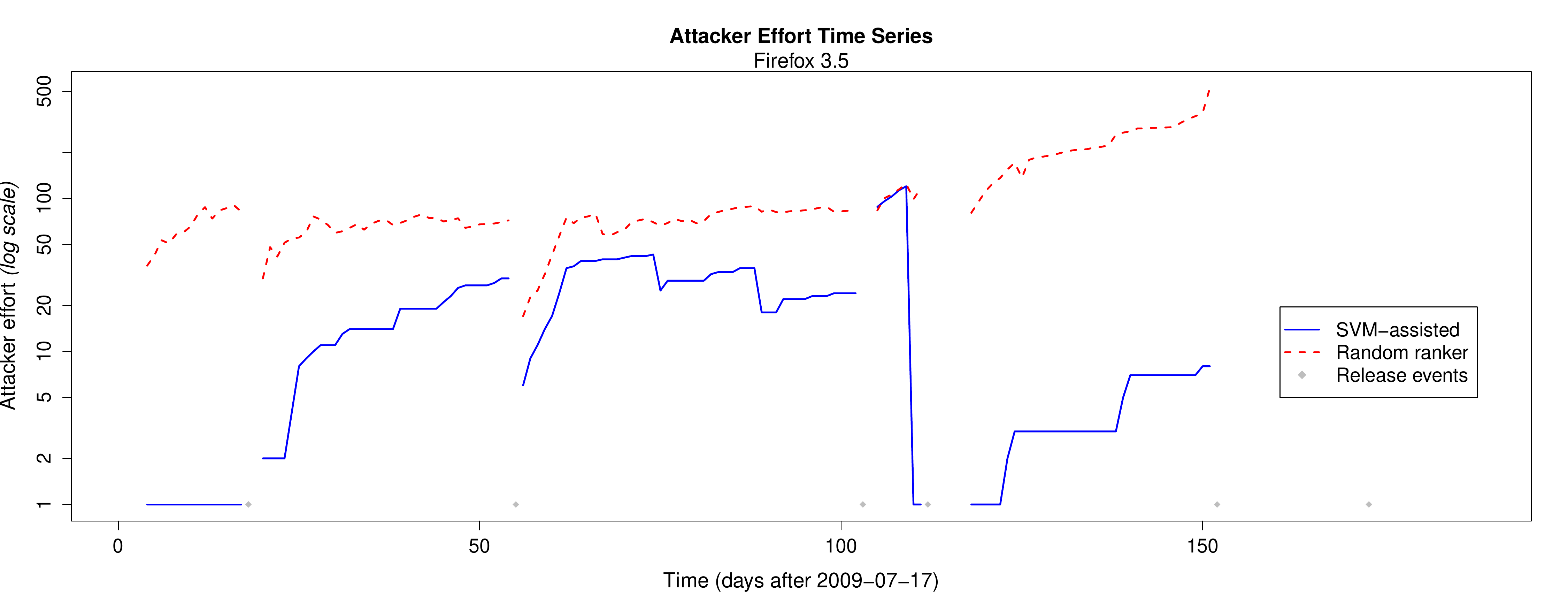}
\caption{SVM-assisted and random ranker efforts for
finding Firefox 3.5 vulnerabilities.}
\label{fig:ff3.5-series}
\end{minipage}
\end{center}
\end{figure*}
\begin{figure*}[t]
\begin{center}
\begin{minipage}[t]{.485\textwidth}
\centering
\includegraphics[width=1\textwidth]{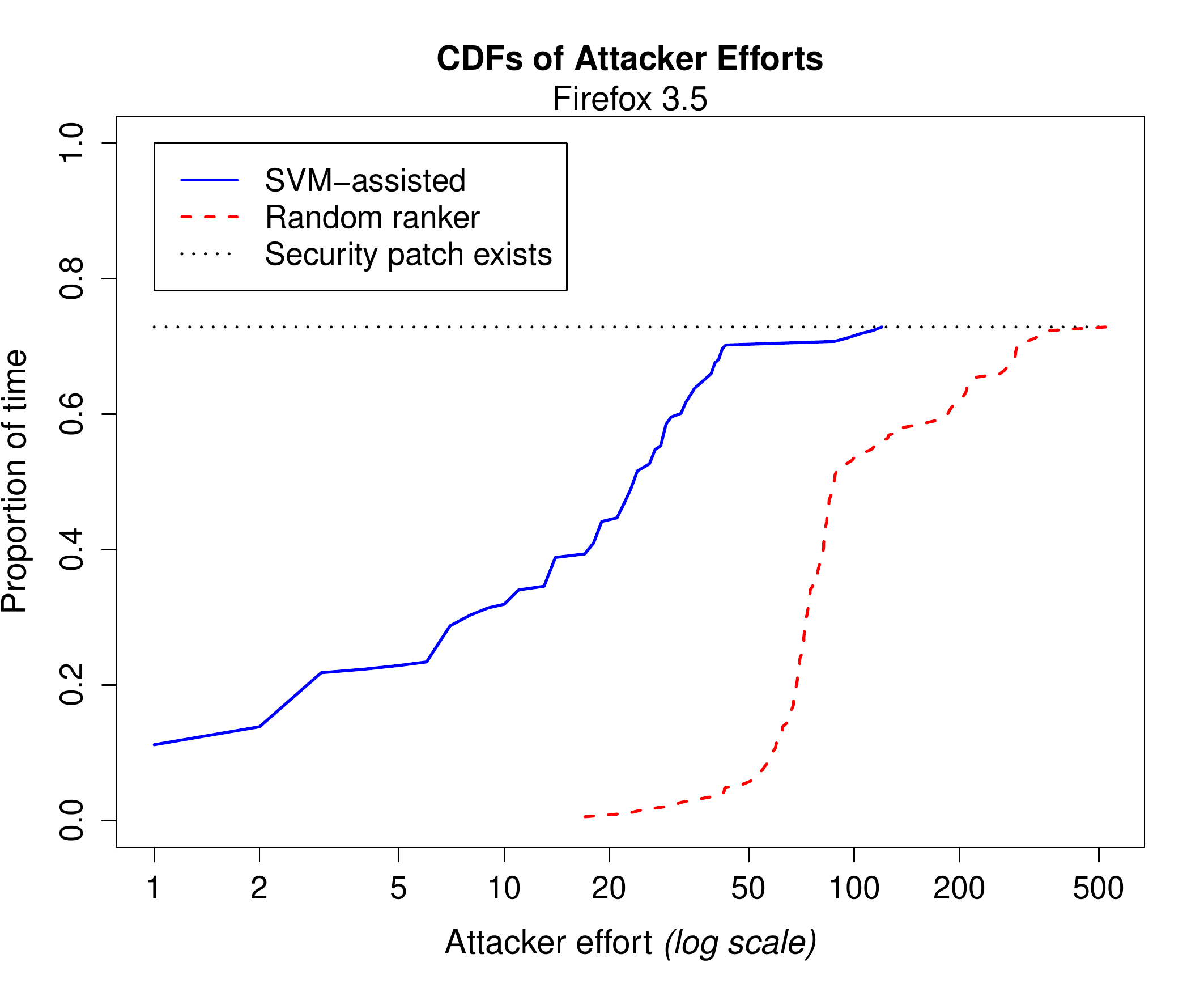}
\caption{The CDFs of the SVM-assisted and random ranker attacker efforts, for Firefox 3.5.}
\label{fig:ff3.5-cdf}
\end{minipage}\hfill
\begin{minipage}[t]{.485\textwidth}
\centering
\includegraphics[width=1\textwidth]{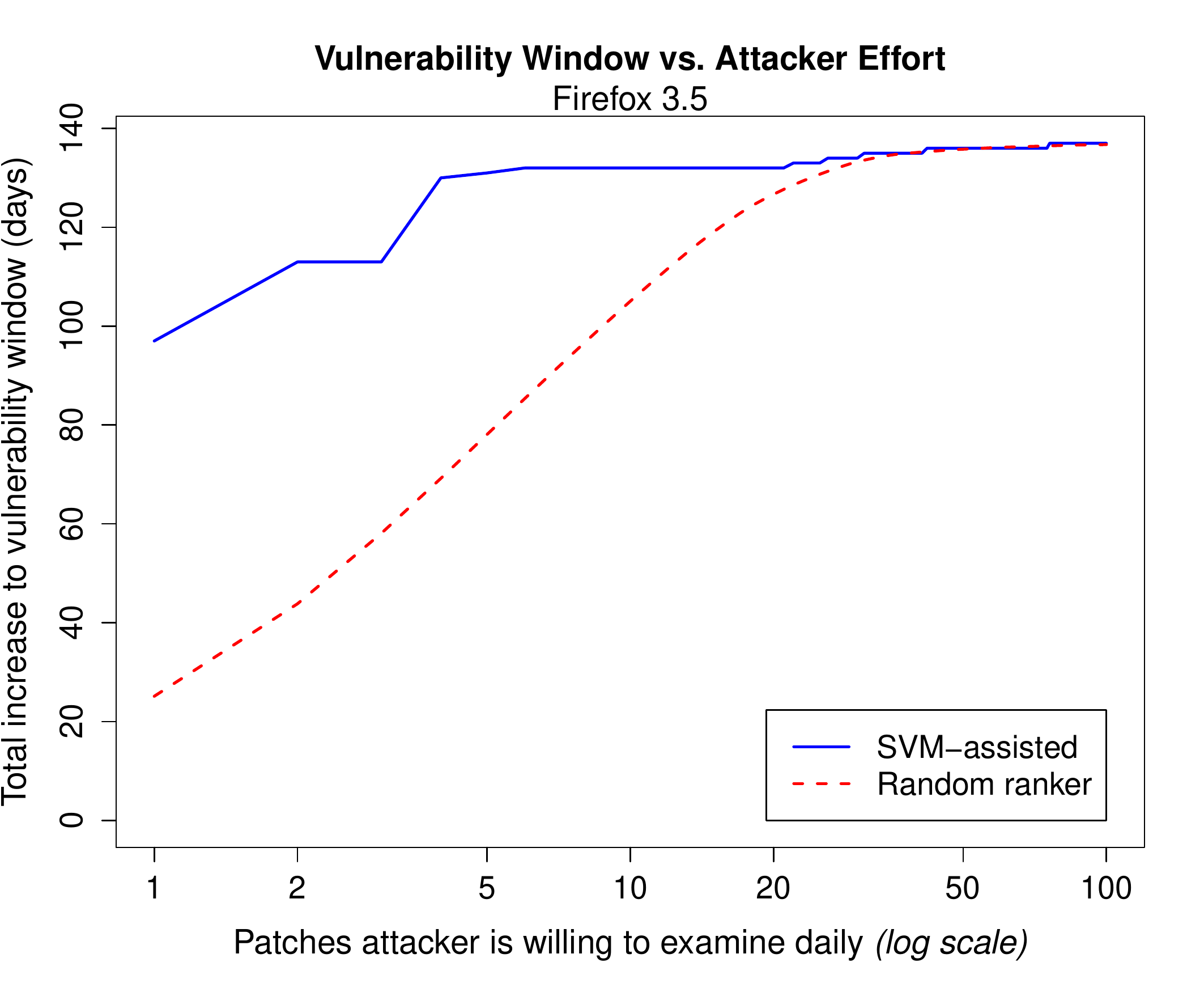}
\caption{Increase to the total window of vulnerability achieved for varying levels of daily attacker effort, for Firefox 3.5.}
\label{fig:ff3.5-vuln-window}
\end{minipage}
\end{center}
\end{figure*}

\subsection{Analysis Repeatability Over Independent Periods of Time}\label{subsec:ff3.5}

In the previous section we explore how an attacker can find vulnerabilities 
over the lifetime of a major release of a large open-source project. It is
natural to ask: how repeatable are these results over subsequent releases?
As a first step towards answering this question, we repeat our analysis on
the complete life-cycle of Firefox 3.5.

While the Firefox 3.5 patch volumes
correspond to roughly half those of the year-long period of active development
on Firefox 3, it is possible that the patches' metadata may have changed subtly,
resulting in significant differences in SVM-assisted ranker performance. Changes
to contributing authors, functions of top-level directories, diff sizes or 
other side-effects of changes to coding policies, time of day or day of week
when patches tend to be landed, could each contribute to changes to the
attacker's performance. Given the similar rates of patch landings, one can
expect the random ranker's performance to be generally unchanged.

Figures~\ref{fig:ff3.5-series}--\ref{fig:ff3.5-vuln-window} depict the
cost-benefit analysis of the SVM-assisted and random rankers searching for
vulnerabilities in Firefox 3.5. It is clear that similar
performance to Firefox 3 is enjoyed. 
The CDFs of attacker effort displayed in Figure~\ref{fig:ff3.5-cdf}
show that while the random ranker's performance is roughly the same as before,
the SVM-assisted ranker's performance at very low effort (1 or 2 patches) is
inferior compared to Firefox 3, while the assisted attacker enjoys much better
performance at low to moderate efforts.

\begin{remark}
The SVM-assisted attacker discovers a security patch in Firefox 3.5 by the third patch examined, for 22\% of the 5.5 month period; by the 20\nth{th} patch the
SVM-assisted attacker finds a security patch for 50\% of the period. By contrast the random ranker must examine $69.1$ or $95$ patches in expectation to find a security patch for these proportions of the 5.5 month period.
\end{remark}

In a similar vein, the increase to the
window of vulnerability achieved by the random ranker is comparable between
Firefox 3 and 3.5 (correcting for the differences in release lifetimes), while
the SVM-assisted attacker achieves superior performance
(\cf\ Figure~\ref{fig:ff3.5-vuln-window}).

\begin{remark}
By examining one or two patches daily, the SVM-assisted ranker increases the window of vulnerability (in aggregate) by 97 or 113 days total (representing increases to the base vulnerability window of factors of 5.8 and 6.7 respectively). By contrast the random ranker achieves increases of 25.1 or 43.8 days total under the same budgets.
\end{remark}

We may conclude from these results that \emph{the presented attacks on
Firefox 3 are repeatable for Firefox 3.5}, and we expect our analysis to extend
to other major releases of Firefox and major open-source projects other than
Firefox.

\ifthenelse{\isundefined{\fullpaper}}{}{
\subsection{Feature Analysis Redux: Measuring the Effect of Obfuscation}\label{subsec:wrapper}

\begin{figure*}[t]
\begin{center}
\begin{minipage}[t]{.485\textwidth}
\centering
\includegraphics[width=1\textwidth]{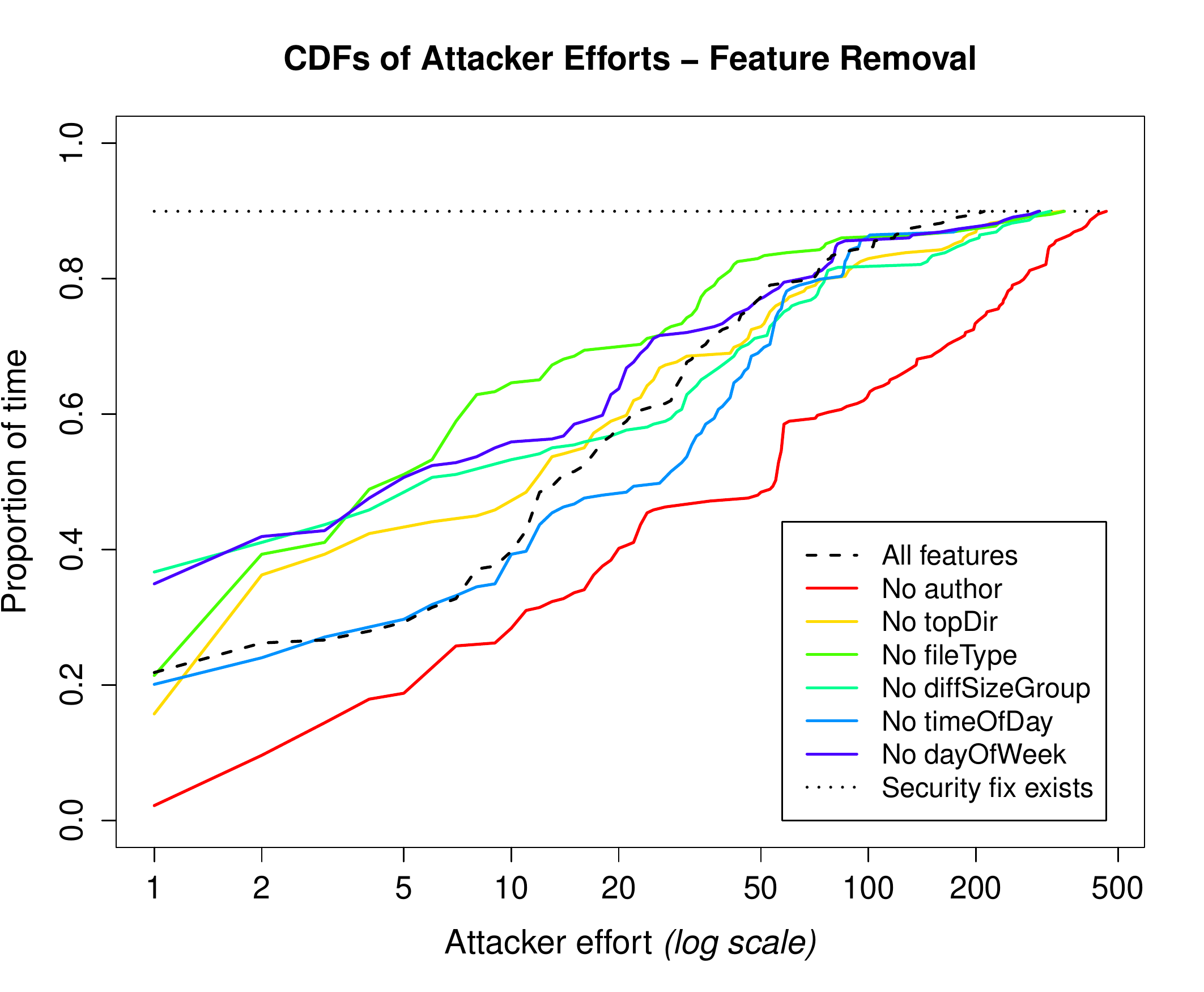}
\caption{The effect of removing individual features on the SVM-assisted attacker effort CDFs, from 11/13/2008 onwards.}
\label{fig:rm-cdf}
\end{minipage}\hfill
\begin{minipage}[t]{.485\textwidth}
\centering
\includegraphics[width=1\textwidth]{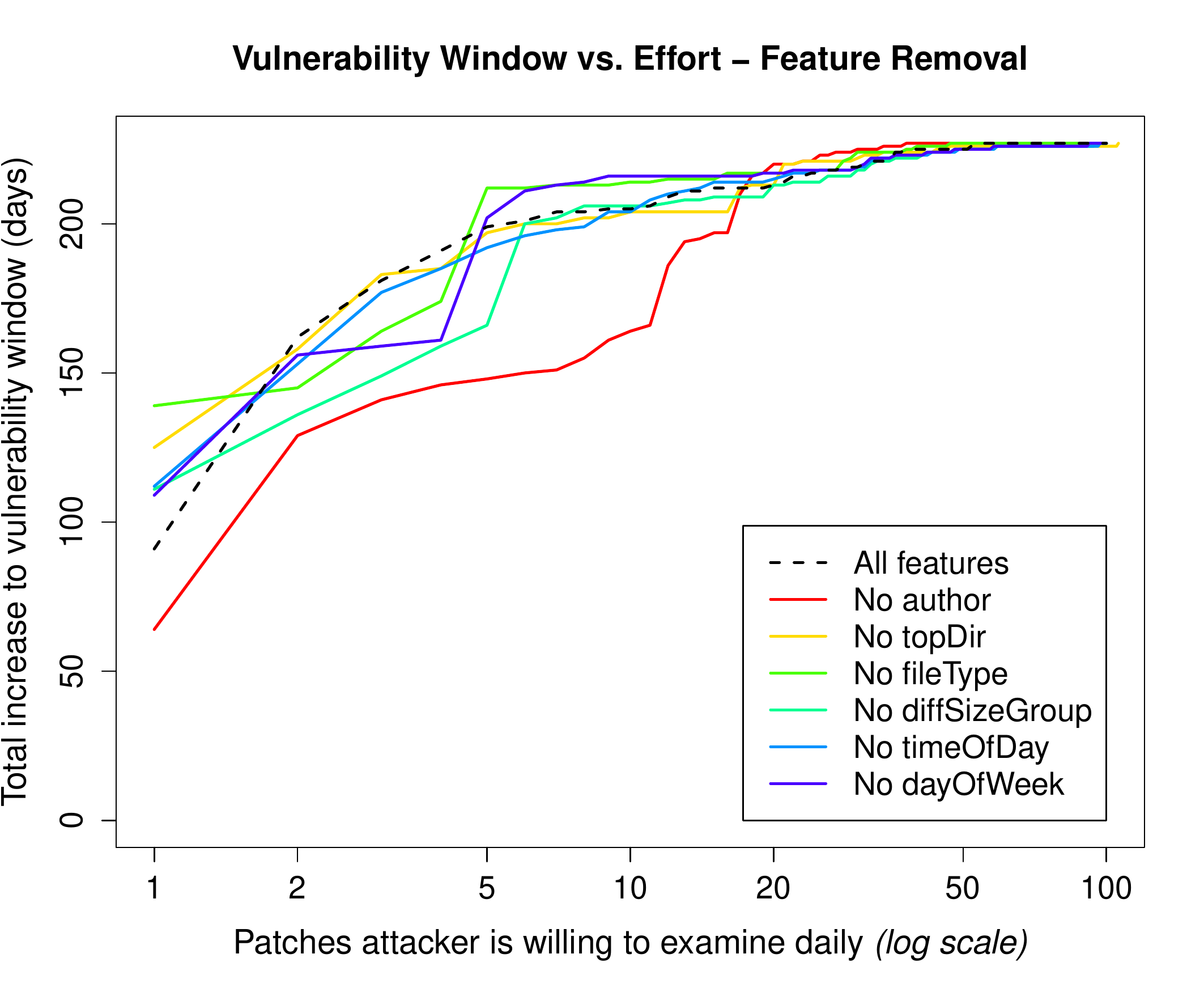}
\caption{The effect of removing features on the SVM-assisted increase to the vulnerability window, from 11/13/2008 onwards.}
\label{fig:rm-vuln-window}
\end{minipage}
\end{center}
\end{figure*}

In Section~\ref{subsec:results-features} we perform a filter-based feature analysis for discriminating between security patches and non-security patches. In this section we ask: what is the effect of obfuscating individual features? We answer this question through a wrapper-based feature analysis in which we perform the same simulation of an SVM-assisted ranker as above, but now with one feature removed.

Figure~\ref{fig:rm-cdf} depicts the attacker effort CDFs for the SVM-assisted ranker when trained with all features, and trained with either the author, top directory, file type, time of day, day of week, or the set of diff size features removed. We remove the number of characters in the diff, number of lines in the diff, number of files in the diff, and file size simultaneously, since we observed no difference when only one of these features was removed. A plausible explanation for this invariance would be high correlation among these features. Removing the author feature has the most negative impact on the attacker effort CDF, reducing the proportion by 0.12 on average over attacker efforts in $[1,464]$. That is, on average over attacker efforts for 12\% of the 8 month period a security patch is found by the SVM-assisted ranker trained with the author feature while the attacker without access to patch author information find no security patch. Removing the time of day has no significant effect; and removing the top directory, day of week, or file type have increasingly positive impacts on the overall attacker effort CDF. Despite \LS's use of cross-validation for tuning the SVM's parameters, the positive improvements point to overfitting which could be a product of the high dimensionality of the learning problem together with a very small sample of security patches: as noted above, our goal is merely to lower bound the performance of an attacker assisted by machine learning.

The increase to the window of vulnerability achieved by an SVM-assisted attacker without access to individual features (or the group of diff size features) is
shown in Figure~\ref{fig:rm-vuln-window}. For some attacker efforts the increase is greater without certain features, but overall we see a more negative effect overall. The greatest negative effect is observed when removing the author feature: the increase in window size is only $8$ days less on average over attacker efforts in $[1,55]$ than when the author feature is included. 

We thus draw the following conclusion, which agrees with the filter-based feature analysis presented in Section~\ref{subsec:results-features}.

\begin{remark}
Obfuscating the patch author has the greatest negative impact on the SVM-assisted ranker's performance, relative to obfuscating other features individually. However the magnitude of impact is negligible.
\end{remark}

As noted above, even if the impact of obfuscating patch authors were greater, doing so would violate the Mozilla Committer's Agreement.
} 

\section{Improving the Security Life-Cycle} \label{sec:secure-lifecyle}

In this section, we explore ways in which open-source projects can avoid
information leaks in their security life-cycle.
\ifthenelse{\isundefined{\fullpaper}}{The feature analysis of Section~\ref{subsec:results-features} and results
(not shown) of removing individual features from the SVM's analysis,
prove that obfuscating the patch author has the greatest individual
negative impact on the SVM-assisted ranker's performance, but that 
this impact is negligible.}{} 
Instead of attempting to
obfuscate metadata from would-be attackers, we
recommend that open-source developers land vulnerability fixes in a ``private''
repository and use a set of trusted testers to ensure the quality of releases.

\subsection{Workflow}

A natural reaction to our experiments is to attempt to plug the information
leaks by obfuscating patches. However, we argue that this approach does not
scale well enough to prevent a sophisticated attacker from detecting security
patches before announcement because an attacker can use standard machine
learning techniques to aggregate information from a number of weak indicators.
In general, it is difficult to predict how such a ``cat-and-mouse'' game would
play out, but, in this case, the attacker appears to have significant
advantage over the defender.

Instead of trying to plug each information leak individually, we recommend
re-organizing the vulnerability life-cycle to prevent information about
vulnerabilities from flowing to the public (regardless of how well the
information is obfuscated). Instead of landing security patches in the public
\MC\ repository first, we propose landing them in a private release branch.
This release branch can then be made public (and the security patches merged
into the public repository) on the day the patch is deployed to users. This
workflow reverses the usual integration path by merging security fixes from
the release branch to \MC\ instead of from \MC\ to the release branch.

\subsection{Quality Assurance}

The main cost of landing security patches later is that the patches receive
less testing before release. When the Firefox developers land security patches
in \MC, those patches are tested by a large number of users who run nightly
builds of Firefox. If a security patch causes a regression (for example, a
crash), these users can report the issue to the Firefox developers before the
patch is deployed to all users. The Firefox developers can then iterate on the
patch and improve the quality of security updates (thereby making it less
costly for users to apply security updates as soon as they are available).

Instead of having the public at large test security updates prior to release,
we recommend that testing be limited to a set of trusted testers. Ideally,
this set of trusted testers would be vetted by members of the security team
and potentially sign a non-disclosure agreement regarding the contents of
security updates. The size of the trusted tester pool is a trade-off between
test coverage and the ease with which an attacker can infiltrate the pool,
which is a risk management decision.

\subsection{Residual Risks}

There are two residual risks with this approach. First, the bug report itself
still leaks some amount of information because the bug is assigned a
sequential bug number that the attacker can probe to determine when a security
bug was filed. This information leak seems fairly innocuous. Second, the
process leaks information about security fixes on the day the patch becomes
available. This leak is problematic because not all users are updated
instantaneously~\cite{Duebendorfer}. However, disclosing the source code contained in each release
is required by many open-source licenses. As a practical matter, source
patches are easier to analyze than binary-only patches, but attackers can
reverse engineer vulnerabilities from binaries alone~\cite{autoexploit}. One
way to mitigate this risk is to update all users as quickly as
possible~\cite{Duebendorfer}.

\section{Conclusions}\label{sec:conclusions}

Landing security patches in public source code repositories significantly
increases the window of vulnerability of open-source projects. Even though
security patches are landed amid a cacophony of non-security patches, we show that an
attacker can exploit patch description fields linking to bug reports to immediately find a
security patch on almost any day of the year. If patch descriptions are obfuscated,
off-the-shelf machine learning can find security patches from patch metadata such as
author. By examining a nominal number of patches daily, these attackers increase the
total window of vulnerability for Firefox by factors of 6 to 10 over the baseline window due
to deployment latency.

A natural reaction to these findings is to obfuscate more features in an
attempt to make the security patches harder to identify. However, our analysis
shows that no single feature contains much information about whether a patch
fixes a vulnerability; and even if all patch metadata is obfuscated, a random ranker can effectively increase the total window of vulnerability by a factor of 4.
Instead of obfuscating patch metadata, we recommend changing
the security life-cycle of open-source projects to avoid landing security
patches in public repositories. We suggest landing these fixes in private
repositories and having a pool of trusted testers test security updates.

Our recommendations reduce the openness of open-source projects by withholding
some patches from the community until the project is ready to release those
patches to end users. However, open-source projects already recognize the need
to withhold some security-sensitive information from the community (as
evidenced by these projects limiting access to security bugs to a vetted
security group). In a broad view, limiting access to the security patches
themselves prior to release is a small price to pay to significantly reduce the window of
vulnerability for open-source software.

 \section*{Acknowledgments}
 
 We would like to thank Pongsin Poosankam, Wil Robertson, Aleksandr Simma, and
 Daniel Veditz for their helpful comments and assistance. We gratefully
 acknowledge the support of the NSF through grant DMS-0707060, and the
 support of the Siebel Scholars Foundation.



\bibliographystyle{IEEEtran}
\bibliography{sources}
%

\ifthenelse{\isundefined{\fullpaper}}{}{
\appendix

\subsection{Proofs for the Random Ranker}\label{app:random}

Here we derive expressions for the random ranker's expected cost (attacker
effort) and benefit (increase to vulnerability window).

\subsubsection{Proof of Random Ranker Expected Effort}\label{app:random-effort}

If the ranker's sampling were performed with replacement, then the
distribution of attacker effort $X$ would be geometric with known expectation.
Without replacement, if there are $n$ patches in the pool, $n_s$ of
which fix vulnerabilities, $X$ has probability mass
\begin{eqnarray}
&& \Pr{X=x} \nonumber \\
&=& \frac{(n-n_s)!}{(n-n_s-x+1)!} \cdot \frac{(n-x+1)!}{n!} \cdot \frac{n_s}{n-x+1} \label{eq:random-prob-part} \\ 
&=& \begin{pmatrix}n-x\\ n_s - 1\end{pmatrix} \begin{pmatrix}n\\ n_s\end{pmatrix}^{-1}\enspace, \label{eq:random-prob}
\end{eqnarray}
for $x\in\{1,\ldots,n-n_s+1\}$ and zero otherwise. The second equality follows
from some simple algebra. The first equality is derived as follows. The
probability of the first draw being a non-security patch is the number of 
non-security patches over the number of patches or $(n-n_s)/n$. Conditioned on
the first patch not fixing a vulnerability, the second draw has probability
$(n-n_s-1)/(n-1)$ of being non-security related since one fewer patch is in the
pool (which is, in particular a non-security patch). This process continues with
the $k\nth{th}$ draw having (conditional) probability $(n-n_s-k+1)/(n-k+1)$
of being a non-security patch. After drawing $k$ non-security patches, the
probability of selecting a patch that fixes a vulnerability is $n_s/(n-k)$. 
Equation~\eref{eq:random-prob-part} follows by chaining these conditional
probabilities.

With $X$'s probability mass in hand, the expectation can be efficiently 
computed for any moderate $(n,n_s)$ pair by summing
Equation~\eref{eq:random-prob}.

\subsubsection{Random Ranker's Expected Increase to the Window of Vulnerability}\label{app:random-window}

We begin by constructing the distribution of the first day an undisclosed vulnerability fix is found after a security update, when the random ranker is constrained to a budget $b$ of patches daily, and never re-examines patches. Let $n_t$ and $n_{t,s}$ denote the number of new patches and new vulnerability fixes landed on day $t\in\mathbb{N}$. Let random variable $X^n_{n_s}$ be the attacker effort required to find one of $n_s$ vulnerability fixes out of a pool of $n$ patches as described above. Finally, let $A_t$ be the event that the first vulnerability fix is found on day $t\in\mathbb{N}$. Then trivially
\begin{eqnarray*}
\Pr{A_1} &=& \Pr{X^{n_1}_{n_{1,s}}\leq b} \enspace.
\end{eqnarray*}
Now we may condition on $\neg A_1$ to express the probability of $A_2$ occurring: if $A_1$ does not occur then $b$ non-security-related patches are removed from the pool, so that the pool consists of $n_{1,s}+n_{1,s}$ vulnerability fixes and $n_1+n_2-b$ patches total. The conditional probability of $A_2$ given $\neg A_2$ is then the probability of $\{X^{n_1+n_2-b}_{n_{1,s}+n_{1,s}}\leq b\}$. By induction we can continue to exploit this conditional independence to yield for all $t>0$
\begin{eqnarray}
&& \Pr{A_{t+1}\mid \neg A_1 \cap\ldots\cap \neg A_t} \\ \nonumber
&=& \Pr{X^{\left(\sum_{i=1}^{t+1}n_i\right) - t b}_{\sum_{i=1}^{t+1} n_{i,s}}\leq b} \label{eq:random-window-inductive}\enspace.
\end{eqnarray}
The RHS of this expression is easily calculated by summing Equation~\eqref{eq:random-prob} over $x\in[b]$. The unconditional probability distribution now follows from the mutual exclusivity of the $A_t$ and the chain rule of probability
\begin{eqnarray*}
&& \Pr{A_{t+1}} \\
&=& \Pr{A_{t+1}\cap \neg A_t \cap \ldots \cap \neg A_1} \nonumber \\
&=& \Pr{A_{t+1}\ \left|\ \bigcap_{s=1}^t \neg A_s \right.} \prod_{i=1}^{t} \Pr{\neg A_i\ \left|\ \bigcap_{s=1}^{i-1}  \neg A_s\right.} \nonumber \\
&=& \Pr{X^{\left(\sum_{i=1}^{t+1}n_i\right) - t b}_{\sum_{i=1}^{t+1} n_{i,s}}\leq b} \\
&& \ \ \cdot\ \prod_{j=1}^t \left(1 - \Pr{X^{\left(\sum_{i=1}^j n_i\right) - (j-1) b}_{\sum_{i=1}^j n_{i,s}}\leq b}\right)\enspace.
\end{eqnarray*}

Thus we need only compute the expression in Equation~\eqref{eq:random-window-inductive} once for each $t\in[N]$, where $N$ is the number of days until the next security update. From these conditional probabilities we can efficiently calculate the unconditional $\Pr{A_t}$ for each $t\in[N]$.
Noting that $A_t$ implies an increase of $Y=N-t+1$ to the window of vulnerability, the expected increase is
\begin{eqnarray}
\Exp{Y} &=& \sum_{t=1}^N (N-t+1)\Pr{A_t} \enspace. \label{eq:exp-window}
\end{eqnarray}

\begin{remark}
Notice that there can be a non-trivial probability that no vulnerability fix will be found by the random ranker in the $N$ day period. This probability is simply $1-\sum_{t=1}^N \Pr{A_t}$. On typical inter-update periods this probability can be higher than 0.5 for budgets $\approx 1$. This fact serves to reduce the expected increase to the window of vulnerability, particularly for small budgets.
\end{remark}

\begin{remark}
The astute reader will notice that we removed $b$ non-security-related patches from the pool on all days we do not find a vulnerability fix, irrespective of whether $b$ or more such patches are present. We have assumed that $n$ is large for simplicity of exposition. Once $n$ drops to $n_s+b$ or lower, we remove all non-security-related patches upon failing to find a vulnerability fix. On the next day, the probability of finding a vulnerability fix is unity. The probabilities of finding vulnerability fixes on subsequent days are thus zero. Thus as $b$ increases, the distribution becomes more and more concentrated at the start of the inter-update period as we would expect. Finally, if no vulnerability fixes are present in the pool on a particular day, then the probability of finding such a patch is trivially zero.
\end{remark}

\subsection{Feature Analysis} \label{sec:infogain}

The information theoretic quantity known as the 
\term{information gain} measures how well a feature separates a set of 
training data, and is popular in information retrieval and in machine learning 
within the ID3 and C4.5 decision tree learning algorithms~\cite{MitchellBook}.

For training set $S$ and nominal feature $F$ taking discrete values in 
$\mathcal{X}_F$, the information gain is defined as
\begin{eqnarray}
 & & \Gain{S,F} \nonumber  \\
&=& \Ent{S_\ell} - \sum_{x\in\mathcal{X}_F}\frac{|S_{\ell,x}|}{|S|}\Ent{S_{\ell,x}}\enspace, \label{eq:info-gain}
\end{eqnarray}
where $S_\ell$ denotes the multiset of $S$'s example binary labels, $S_{\ell,x}$
denotes the subset of these labels for examples with feature $F$ value $x$, and
for multiset $T$ taking possible values in $\mathcal{X}$ we have the usual definition of $\Ent{T} = -\sum_{x\in\mathcal{X}}\frac{|T_x|}{|T|}\log_2\frac{|T_x|}{|T|}$. The first term of the information gain, the entropy of the training set,
corresponds to the impurity of the examples' labels. A pure set with only one
repeated label has zero entropy, while a set having half positive examples and
half negative examples has a maximum entropy of one. The information gain's
second term corresponds to the expected entropy of the training set conditioned
on the value of feature $F$. Thus a feature having a \emph{high} information
gain corresponds to a large drop in entropy, meaning that splitting on that
feature resulting in a partition of the training set into subsets of like
labels. A \emph{low} (necessarily non-negative) information gain corresponds to
a feature that is not predictive of class label.

Two issues require modification of the basic information gain before use in
practice~\cite{MitchellBook}. The first is that nominal features $F$ with
large numbers
of discrete values $|\mathcal{X}_F|$ tend to have artificially inflated
information gains (being as high as $\log_2|\mathcal{X}_F|$) since splitting
on such features can lead to numerous small partitions of the training set with
trivially pure labels. An example is the author feature, which has close to 500
values. In such cases it is common practice to correct for this 
artificial inflation by using the \term{information gain ratio}~\cite{dtrees}
as defined
below. We use $S_F$ to denote the multiset of examples' feature $F$ values in
the ratio's denominator, which is known as the \term{split information}.
\begin{eqnarray}
\GainRatio{S,F} &=& \frac{\Gain{S,F}}{\Ent{S_F}}\enspace.\label{eq:info-gain-ratio}
\end{eqnarray}

The second issue comes from taking the idea of many-valued nominal features to
the extreme: continuous features such as the diff length (of which there are
7,572 unique values out of 14,541 patches in our dataset) and the file size
(which enjoys 12,795 unique values) are analyzed by forming a \term{virtual
binary feature} for each possible threshold on the feature. The information gain
(ratio) of a continuous feature is defined as the maximum information gain
(ratio) of any induced virtual binary feature~\cite{dtrees-thesis}.

}

\end{document}